\documentclass[a4paper,11pt]{article}
\pdfoutput=1
\usepackage[utf8x]{inputenc}
\usepackage{jheppub}
\usepackage{epsf}
\usepackage{color}
\usepackage{subfigure}
\usepackage{ulem}
\usepackage{slashed}
\newcommand{\beq}{\begin{eqnarray}}

\newcommand{\bea}{\begin{eqnarray}}
\newcommand{\eea}{\end{eqnarray}}
\newcommand{\eq}[1]{Eq.~(\ref{#1})}

\newcommand{\nn}{\nonumber}

\def\stw{s_{\theta_W}}
\def\ctw{c_{\theta_W}}
\def\ttw{t_{\theta_W}}
\def\lra#1{\overset{\text{\scriptsize$\leftrightarrow$}}{#1}}

\makeatletter
\def\section{\@startsection {section}{1}{\z@}{-3.5ex plus -1ex minus
 -.2ex}{2.3ex plus .2ex}{\large\bf}}
\def\subsection{\@startsection{subsection}{2}{\z@}{-3.25ex plus -1ex
minus -.2ex}{1.5ex plus .2ex}{\normalsize\bf}}
\makeatother
\makeatletter

\@addtoreset{equation}{section}

\makeatother

\preprint{IPPP/19/93}
\begin{document}

\title{ Towards the ultimate differential SMEFT analysis}
\author{Shankha Banerjee, Rick S. Gupta, Joey Y. Reiness, Satyajit Seth and Michael Spannowsky}
\affiliation{Institute for Particle Physics Phenomenology,
Durham University, South Road, Durham, DH1 3LE}

\date{\today}

\abstract{We obtain SMEFT bounds using an approach that utilises the complete multi-dimensional differential information of a process. This approach is based on the fact that at a given EFT order, the full angular distribution in the most important electroweak processes can be expressed as a sum of a fixed number of basis functions. The coefficients of these basis functions - the so-called angular moments - and their energy dependance, thus form an ideal set of experimental observables that encapsulates the complete multi-dimensional differential information of the process. This approach is generic and the observables constructed allow to avoid blind directions in the SMEFT parameter space. While this method is applicable  to many of the important electroweak processes, as a first example we study the $pp \to V(\ell\ell)h(bb)$ process ($V \equiv Z/W^{\pm}, \; \ell\ell \equiv \ell^+\ell^-/\ell^\pm\nu$),  including QCD NLO effects, differentially. We show that  using the full differential data in this way plays a crucial role in simultaneously and maximally constraining the different vertex structures of the Higgs coupling to gauge bosons. In particular, our method yields  bounds on the $h V_{\mu \nu}V^{\mu \nu}$, $h V_{\mu \nu}\tilde{V}^{\mu \nu}$ and $h Vff$ ($ff \equiv f\bar{f}/f\bar{f}'$) couplings, stronger than projected bounds reported in any other process. This matrix-element-based method can provide a transparent alternative to complement machine learning techniques that also aim to disentangle correlations in the SMEFT parameter space.}

\maketitle

\section{Introduction}

The data being collected by the LHC is  the first record of interactions of the Higgs and other Standard Model (SM) particles at the sub-attometre (multi TeV)  scale. As long as beyond SM (BSM) physics is significantly heavier than the mass of electroweak particles, these interactions can be described in a model independent way by the Standard Model Effective Field Theory (SMEFT) Lagrangian. The SMEFT Lagrangian is thus a statement of the laws of nature at the most fundamental scale ever probed. The measurement of (or constraints on) the SMEFT parameters~\cite{Buchmuller:1985jz, Giudice:2007fh, Grzadkowski:2010es, Gupta:2011be,Gupta:2012mi,Banerjee:2012xc, Gupta:2012fy, Banerjee:2013apa, Gupta:2013zza,Elias-Miro:2013eta, Contino:2013kra, Falkowski:2014tna,Englert:2014cva, Gupta:2014rxa, Amar:2014fpa, Buschmann:2014sia, Craig:2014una, Ellis:2014dva, Ellis:2014jta, Banerjee:2015bla, Englert:2015hrx, Ghosh:2015gpa, Cohen:2016bsd, Ge:2016zro, Contino:2016jqw, Biekotter:2016ecg, deBlas:2016ojx, Denizli:2017pyu, Barklow:2017suo, Brivio:2017vri, Barklow:2017awn, Khanpour:2017cfq, Englert:2017aqb, panico, Banerjee:2018bio, Grojean:2018dqj,Biekotter:2018rhp, Goncalves:2018ptp,Freitas:2019hbk,Banerjee:2019pks} may well turn out to be the main legacy of the LHC after the Higgs discovery.

It is thus of great importance to maximally exploit all the data that the LHC would provide us. To constrain the SMEFT Lagrangian, it is especially important to extract the full multi-dimensional differential information available  in a process.  This is because the effect of new vertex structures arising at the dimension-6 (D6) level is often more pronounced in certain regions of the phase space, the most common example being the growth of EFT rates at high energies. A more subtle example is that of operators whose contributions do not interfere with the SM amplitude at the inclusive level~\cite{nonint}. These operators can generate large excesses differentially~\cite{Banerjee:2019pks, Hagiwara:1986vm, azatov, panico, azatov2, Azatov:2019xxn} in certain regions of the phase space, which are cancelled by corresponding deficits in other   regions. These effects can, therefore,  get lost unless a sophisticated study is carried out to isolate these phase space regions. As discussed in Ref.~\cite{Banerjee:2019pks}, and as we will also see in this work, sometimes in order to resurrect these interference terms one has to go even  beyond  differential  distributions with respect to a single variable and  use multidimensional  distributions.   More generally, using the full differential information enlarges the list of observables and lifts flat directions in EFT space that can otherwise remain unconstrained.  In order to   optimally  reconstruct the SMEFT lagrangian, it is thus  essential to  systematically and completely  extract \textit{all} the available differential information.

In the way experimental measurements are communicated, there is a large reduction in differential information, as often only a few intuitively chosen distributions are presented. To estimate this, consider a  three body final state where the phase space in the center of mass frame can be completely described by four variables: an energy variable and three angles.  For a given energy, taking for instance 10 bins for each of the angular variables results in 1000 units of data to capture the entire information contained in this process, at this level of experimental precision. However, often individual angles are analysed in isolation and the correlations contained in the full set of data are projected onto only 30 units of data, \textit{i.e.}, 10 for each angle, resulting in a loss of accessible information to search for new physics contributions.

Interestingly, for many important processes the 1000 units of data, contain redundant information. We argue, that with an understanding of the underlying theoretical structure of process the number of physical quantities required to completely characterise the full differential distribution can be drastically reduced. The main fact that we will utilise in this work is that, for some of the most important processes in Higgs and electroweak physics, the full angular distribution at a given energy can be expressed as a sum of a fixed number of basis functions as long as we limit ourselves to a certain order in the EFT expansion.  The reason for this is that only a finite number of helicity amplitudes get corrections up to the  given EFT order, see for instance Ref.~\cite{Bellazzini:2018paj, Melia}.  The coefficients of these basis functions, the so called angular moments~\cite{Dunietz:1990cj,  Dighe:1998vk, james, Beaujean:2015xea}, and their energy dependance, thus, contain the full differential information available in a process. The effect of EFT operators on differential distributions can therefore be summarised by their contribution to these angular moments. As such angular moments can be used to construct any possible differential distribution, an analysis utilising them has the potential to reach maximal sensitivity in probing EFT coefficients. 

While similar approaches have been used for some isolated studies in Higgs and  flavour physics~\cite{Hagiwara:1986vm, Dighe:1998vk, oo1,oo2,oo3,Godbole:2007cn, james, Godbole:2013lna, Godbole:2014cfa, Beaujean:2015xea, Gratrex:2015hna}, we believe the suitability of these techniques in globally constraining the SMEFT lagrangian have not been  sufficiently recognised. 
 
These methods would complement other techniques that aim to employ a maximum-information approach, e.g. the matrix element method~\cite{matrix, Gainer:2013iya, Soper:2014rya,Brehmer:2016nyr, Brehmer:2017lrt,Brehmer:2018hga,Brehmer:2019gmn,Brehmer:2019xox,Prestel:2019neg} or machine learning techniques that have recently gained popularity~\cite{machine1, machine2,Englert:2018cfo, DAgnolo:2018cun, machine4}. One advantage of this approach over other multivariate  techniques is its more physical and transparent nature. The angular moments described above can be directly related to physical experimental quantities, \textit{e.g.} they have well defined symmetry properties, than the abstract neural network outputs used in machine learning approaches. Another important distinction of the methods proposed here  from some multivariate  approaches like the matrix element method, is that the process of extraction of the angular moments is hypothesis-independent; for instance it would be independent of our assumptions about whether electroweak symmetry is linearly or non-linearly realised.

In this work we will show how these angular moments can be extracted and mapped back to the EFT  lagrangian. While in this study  we will focus on Higgs-strahlung at the LHC as a first example, this approach can be extended to all the important Higgs/electroweak production and decay processes, namely weak boson fusion, Higgs decay to weak bosons and diboson production. For the Higgs-strahlung process at the partonic level there are 9 angular moments, although a smaller number of these are measurable at the LHC for the final states that we are interested in. We will see that extracting all the experimentally  available angular moments can simultaneously constrain all the possible $hVV^*/hVff$ ($V \equiv Z/W^{\pm}, ff \equiv f\bar{f}/f\bar{f}'$) tensor structures. An essential prerequisite for our methods to be applicable is that the final angular distributions measured by the experiments should preserve, to a large extent, the initial theoretical form of EFT signal governed by the angular moments. To truly establish the usefulness of our methods, we therefore carry out a detailed and realistic  collider study. In particular we include differentially  QCD NLO effects that can potentially improve partonic contributions to the EFT signal reducing scale uncertainties.  In our final results we find, despite these effects, a marked improvement in sensitivity  compared to existing projections for most of the EFT couplings.

The paper is divided as follows. In Sec.~\ref{eft}, we write the most general Lagrangian for the $pp \to V(\ell \ell)h(b\bar{b})$ at Dimension 6 in SMEFT and list the relevant operators in the Warsaw basis. Sec.~\ref{angmom} is dedicated in deriving the most general angular moments for the $pp \to Vh$ processes in the SMEFT. In Sec.~\ref{moments}, we discuss the  method of moments. In Sec.~\ref{collider}, we detail the collider studies that we undertake for the $pp \to Vh$ processes. Sec.~\ref{analysis} is where we discuss the details of the angular analyses and obtain the bounds on the various couplings. We finally conclude in Sec.~\ref{conclusions}.

\section{The $pp \to V(\ell \ell)h(b\bar{b})$ process  in the Dimension 6 SMEFT}
\label{eft}

We want to study the process $pp \to V(\ell \ell)h(b\bar{b})$ where $\ell \ell$ denotes $\ell^+ \ell^- (\ell^+ \nu, \ell^- \bar{\nu})$ for $V=Z \; (V=W^{\pm})$. The EFT corrections to  $pp \to V(\ell\ell)h(b\bar{b})$ are either due to corrections of the $Vff$,  $hb\bar{b}$ and $hVV/hZ\gamma$ vertices or due to the new $hVff$ contact terms. In the unitary gauge all these corrections are contained in the following Lagrangian ~\cite{Gupta:2014rxa, Pomarol:2014dya}),
\bea
\Delta {\cal L}_6&\supset&    \delta \hat{g}^h_{WW}\, \frac{2 m_W^2}{v}h W^{+\mu}W^-_\mu + \delta \hat{g}^h_{ZZ}\, \frac{2 m_Z^2}{v}h \frac{Z^\mu Z_\mu}{2}+ \delta g^W_{Q}\, (W^+_\mu \bar{u}_L \gamma^\mu d_L+h.c.) 
\nn\\
&+& \delta g^W_{L}\,(W^+_\mu \bar{\nu}_L \gamma^\mu e_L+h.c.)+ g^h_{WL}\,\frac{h}{v}(W^+_\mu \bar{\nu}_L \gamma^\mu e_L+h.c.)\nn\\&+& g^h_{WQ}\,\frac{h}{v}(W^+_\mu \bar{u}_L \gamma^\mu d_L+h.c.)+\sum_f \delta g^Z_{f} Z_\mu \bar{f} \gamma^\mu f +\sum_f g^h_{Zf}\,\frac{h}{v}Z_\mu \bar{f} \gamma^\mu f \nn\\
&+&\kappa_{WW}\,\frac{h}{v} W^{+\mu\nu}W^-_{\mu\nu}+\tilde{\kappa}_{WW}\,\frac{h}{v} W^{+\mu\nu}\tilde{W}^-_{\mu\nu} +\kappa_{ZZ}\,\frac{h}{2v} Z^{\mu\nu}Z_{\mu\nu}\nn\\ &+&\tilde{\kappa}_{ZZ}\,\frac{h}{2v} Z^{\mu\nu}\tilde{Z}_{\mu\nu}+\kappa_{Z\gamma}\,\frac{h}{v} A^{\mu\nu}Z_{\mu\nu}\nonumber
+\tilde{\kappa}_{Z\gamma}\,\frac{h}{v} A^{\mu\nu}\tilde{Z}_{\mu\nu}\,+\delta \hat{g}^h_{b\bar{b}} \frac{\sqrt{2} m_b}{v} h b\bar{b},
\label{anam}
\eea
where for brevity we have  only included the first generation for the couplings involving $W^\pm,Z$ bosons, so that  $f=u_L, d_L, u_R,$ $d_R,e_L, e_R, \nu^e_L$; $F=Q (L)$, the first generation quark (lepton) doublet. We assume that the above Lagrangian is extended to the other generations in a way  such that the couplings $\delta g^{Z,W}_f$ and $g^{h}_{Zf,Wf}$ are flavour diagonal and universal in the interaction basis, allowing us to impose strong constraints on them~\cite{Pomarol:2013zra,Falkowski:2014tna} (this is well motivated theoretically and can be obtained, for instance, by including the leading terms after imposing Minimal Flavour Violation~\cite{DAmbrosio:2002vsn}). If we limit ourselves to only universal corrections, the contact terms  above must be replaced by $h V_\mu \partial_\nu V^{\mu \nu}$ (note that $\partial_\mu h V_\nu  V^{\mu \nu}$ is equivalent to this vertex and the   $h V_{\mu \nu} V^{\mu \nu}$ vertices via integration by parts). The above parametrisation can be used even for non-linearly realised electroweak symmetry (see for \textit{eg.},~\cite{Isidori:2013cga}) and in this case all the above couplings should be thought of as independent.


\begin{table}[t]
\small
\centering
\begin{tabular}{c}
\begin{tabular}{||c|c||}
\hline
&\\
                ${\cal O}_{H\square}=(H^\dagger H) \square (H^\dagger H)$&${\cal O}^{(3)}_{HL}=i H^\dagger \sigma^a \lra{D}_\mu H \bar{L}  \sigma^a \gamma^\mu L$ \\
\rule{0pt}{4ex} ${\cal O}_{HD}=(H^\dagger  {D}_\mu H)^*(H^\dagger  {D}_\mu H)$ &${\cal O}_{HB}= |H|^2 B_{\mu\nu}B^{\mu\nu}$\\
\rule{0pt}{4ex} ${\cal O}_{Hu}=i H^\dagger \lra{D}_\mu H \bar{u}_R  \gamma^\mu u_R$&${\cal O}_{HWB}=  H^\dagger \sigma^a H W^a_{\mu\nu}B^{\mu\nu}$\\
\rule{0pt}{4ex} ${\cal O}_{Hd}=i H^\dagger \lra{D}_\mu H \bar{d}_R  \gamma^\mu d_R$&${\cal O}_{H{W}}= |H|^2 W_{\mu\nu}{W}^{\mu\nu}$\\
\rule{0pt}{4ex} ${\cal O}_{He}=i H^\dagger \lra{D}_\mu H \bar{e}_R  \gamma^\mu e_R$&${\cal O}_{H\tilde{B}}= |H|^2 B_{\mu\nu}\tilde{B}^{\mu\nu}$ \\
\rule{0pt}{4ex} ${\cal O}^{(1)}_{HQ}=i H^\dagger  \lra{D}_\mu H \bar{Q}   \gamma^\mu Q$&${\cal O}_{H\tilde{W}B}=  H^\dagger \sigma^a H W^a_{\mu\nu}\tilde{B}^{\mu\nu}$ \\
\rule{0pt}{4ex} ${\cal O}^{(3)}_{HQ}=i H^\dagger \sigma^a \lra{D}_\mu H \bar{Q}  \sigma^a \gamma^\mu Q$&${\cal O}_{H\tilde{W}}= |H|^2 W^a_{\mu\nu}\tilde{W}^{a \mu\nu}$ \\
\rule{0pt}{4ex} ${\cal O}^{(1)}_{HL}=i H^\dagger  \lra{D}_\mu H \bar{L}   \gamma^\mu L$&${\cal O}_{y_b}=   |H|^2(\bar{Q}_3 H b_R+h.c).$ \\
&\\
\hline
 \end{tabular}
\end{tabular}
\caption{Dimension-6 operators in the Warsaw basis that contribute to the anomalous $hVV^*/hV\bar{f}f$ couplings in \eq{anam}.Other details regarding the notation can be found in~\cite{Grzadkowski:2010es}.}
\label{opers}
\end{table}
 
If electroweak symmetry is  linearly realised, the above vertices arise in the unitary gauge from electroweak invariant operators containing the Higgs doublet. For instance, the operators of the  Warsaw basis~\cite{Grzadkowski:2010es} in Table~\ref{opers},  give the following contributions to these vertices, 
 \bea
 \label{wilson}
 \delta g^W_{f}&=& \frac{g}{\sqrt{2}}\frac{v^2}{\Lambda^2} c^{(3)}_{HF}+\frac{\delta m^2_Z}{m^2_Z}\frac{\sqrt{2}g\ctw^2}{4\stw^2}\nn\\
   g^h_{Wf}&=& \sqrt{2} g\frac{v^2}{\Lambda^2} c^{(3)}_{HF}\nn\\
    \delta \hat{g}^h_{WW}&=&\frac{v^2}{\Lambda^2}\left(c_{H\square}-\frac{c_{HD}}{4}\right)\nn\\
\kappa_{WW}&=&\frac{2 v^2}{ \Lambda^2}c_{HW}\nn\\
  \tilde{\kappa}_{WW}&=&\frac{2 v^2}{ \Lambda^2}c_{H\tilde{W}}\nn\\
   \delta g^Z_{f}&=&-\frac{g Y_f \stw}{\ctw^2}\frac{v^2}{\Lambda^2} c_{WB} -\frac{g}{\ctw}\frac{v^2}{\Lambda^2}(|T_3^f|c^{(1)}_{HF}-T_3^f c^{(3)}_{HF}+(1/2-|T_3^f|)c_{Hf})\nn\\&+&\frac{\delta m^2_Z}{m^2_Z}\frac{g}{2\ctw\stw^2}(T_3 \ctw^2+Y_f \stw^2)\nn\\
  \delta \hat{g}^h_{ZZ}&=&\frac{v^2}{\Lambda^2}\left(c_{H\square}+\frac{c_{HD}}{4}\right)\nn\\
  g^h_{Zf}&=&- \frac{2 g}{\ctw}\frac{v^2}{\Lambda^2}(|T_3^f|c^{(1)}_{HF}-T_3^f c^{(3)}_{HF}+(1/2-|T_3^f|)c_{Hf})\nn\\
  \kappa_{ZZ}&=&\frac{2 v^2}{ \Lambda^2}(\ctw^2 c_{HW}+\stw^2 c_{HB}+ \stw \ctw c_{HWB})\nn\\
  \tilde{\kappa}_{ZZ}&=&\frac{2 v^2}{ \Lambda^2}(\ctw^2 c_{H\tilde{W}}+\stw^2 c_{H\tilde{B}}+ \stw \ctw c_{H\tilde{W}B})\nn\\
   \kappa_{Z\gamma}&=&\frac{v^2}{ \Lambda^2}(2 \ctw \stw(c_{H{W}}- c_{H{B}})+ (\stw^2- \ctw^2) c_{H{W}B})\nn\\
  \tilde{\kappa}_{Z\gamma}&=&\frac{ v^2}{ \Lambda^2}(2\ctw \stw (c_{H\tilde{W}}- c_{H\tilde{B}})+ (\stw^2- \ctw^2) c_{H\tilde{W}B})\nn\\
  \delta \hat{g}^h_{b\bar{b}}&=&- \frac{v^2}{\Lambda^2} \frac{v}{\sqrt{2}m_b}c_{y_b}+\frac{v^2}{\Lambda^2}(c_{H\square}- \frac{c_{HD}}{4}),
 \eea
 where we have used $(m_W, m_Z,\alpha_{em},m_b)$ as our input parameters.  In the  equations for $\delta g^{W,Z}_f$ above, the term,
\bea
\frac{\delta m^2_Z}{m^2_Z}= \frac{v^2}{\Lambda^2}(2 \ttw c_{WB}+\frac{c_{HD}}{2}),
\eea
makes explicit the contribution to the shift in the input parameter, $m_Z$,  due to the above operators.

The $pp \to W^\pm(\ell\nu)h(b\bar{b})$ process directly constrains the couplings $\delta \hat{g}^h_{WW}, \kappa_{WW}$ and $g^h_{WQ}$,  whereas the $pp \to Z(l^+ l^-)h(b\bar{b})$ process constrains the couplings  $ \delta \hat{g}^h_{ZZ}$, a linear combination of  $\kappa_{ZZ}$ and  $\kappa_{Z\gamma}$,  and  the following linear combination of the contact terms~\cite{Banerjee:2018bio}, 
 \bea
g^h_{Z\textbf{p}}=g^h_{Zu_L} -0.76~g^h_{Zd_L}   - 0.45~g^h_{Zu_R} + 0.14~g^h_{Zd_R}  \,.
\label{ghzp}
\eea 
This linear combination arises by summing over the polarisations of the initial quarks as well as including the possibility of both up and down type initial-state quarks weighted by their respective PDF luminosities; the precise  linear combination changes very little with energy.

For the case of linearly realised electroweak symmetry, the $CP$-even couplings involved in $W^\pm h$ production can be correlated to those involved in $Zh$ production using the fact that the same set of operators in Table~\ref{opers} generate all the anomalous couplings as shown in \eq{wilson}.  To derive these correlations we can  trade the 13 $CP$-even Wilson coefficients above for  the 13 independent (pseudo-)observables $\delta \hat{g}^h_{b\bar{b}}$, $\delta g^Z_f$ (7 couplings),  $g^h_{WQ}$, $\delta \hat{g}^h_{WW}$, $\kappa_{WW}$, $\kappa_{Z\gamma}$ and $\kappa_{\gamma \gamma}$, the coefficient of $\frac{h}{2v}A_{\mu \nu} A^{\mu \nu}$~\footnote{This analysis is in the spirit of  Ref.~\cite{Gupta:2014rxa} but with a different choice of primary/independent observables. Indeed, we include in our list the anomalous Higgs couplings, $g^h_{WQ}$ and  $\kappa_{ZZ}$,  rather than  the anomalous triple gauge couplings (TGC) $\delta \kappa_\gamma$ and $\delta g^Z_1$. As we will see, the bounds on the anomalous Higgs couplings are comparable or better than those expected for the TGCs. }. This can be done  using the expressions in \eq{wilson} and the corresponding expression for $\kappa_{\gamma \gamma}$,
\bea
\label{kaa}
\kappa_{\gamma\gamma}=\frac{2 v^2}{ \Lambda^2}(\stw^2 c_{HW}+\ctw^2 c_{HB}- \stw \ctw c_{HWB}).
\eea
The rest of the anomalous couplings can then be expressed as functions of these independent ones; for example we obtain,
\bea
\delta \hat{g}^h_{ZZ}&=&\delta \hat{g}^h_{WW}-\left(\kappa_{WW} - \kappa_{\gamma\gamma} -  \kappa_{Z\gamma}\frac{\ctw}{\stw}\right)\frac{\stw^2}{\ctw^2}+\left(\sqrt{2} \ctw(\delta g^Z_{u_L}-\delta g^Z_{d_L})-g^h_{WQ}\right)\frac{\stw^2}{\sqrt{2} g \ctw^2}\nn\\
\kappa_{ZZ}&=&\frac{1}{\ctw^2} (\kappa_{WW}-2 \ctw \stw\kappa_{Z\gamma}-\stw^2 \kappa_{\gamma \gamma}) \,.
\eea
Some of the couplings on the right-hand side of the above equations can be measured extremely precisely. For instance, the  two couplings, $\kappa_{Z\gamma}$ and $\kappa_{\gamma\gamma}$, can be bounded very strongly (below per-mille level) by measuring the $h\to \gamma \gamma/\gamma Z$ branching ratios~\cite{Pomarol:2013zra, Elias-Miro:2013eta}~\footnote{This might seem surprising, as the branching ratios themselves are not constrained at this level. Recall, however, that the SM  $h\to \gamma \gamma/\gamma Z$ rates are loop suppressed, so that even an ${\cal O}(10 \%)$ uncertainty in the branching ratios translate to per-mille level bounds on these couplings.}. In addition, the $Z$-coupling deviations, $\delta g^Z_f$, are constrained at the per-mille level by LEP data~\cite{Falkowski:2014tna}. As we will see later,  studying $W^\pm h$ production at high energies would allow us to constrain  $g^h_{WQ}$ at the per-mille level. On the other hand, the couplings   $\kappa_{VV}$ and $\delta \hat{g}^h_{VV}$  can be constrained at most at the 1-10$\%$ level. Thus, one can safely ignore the strongly-constrained  couplings  to obtain the direct relationships,
\bea
\label{relation}
\delta \hat{g}^h_{ZZ}&\approx&\delta \hat{g}^h_{WW}-\frac{\stw^2}{\ctw^2} \kappa_{WW},\nn\\
\kappa_{ZZ}&\approx& \frac{\kappa_{WW}}{\ctw^2}\,,
\eea
which hold up to a very good approximation. We will utilise these relationships in order to  combine our results from $W^\pm h$  and $Zh$ modes to  obtain our final bounds on the $CP$-even vertices.

As far as the $CP$-odd couplings are concerned there are 4 of them including those corresponding to $\frac{h}{2v}A_{\mu \nu} \tilde{A}^{\mu \nu}$ and $\frac{h}{2v}A_{\mu \nu} \tilde{Z}^{\mu \nu}$. The latter two couplings are, however, not precisely measurable as in the $CP$-even case. Thus an analog of the above procedure to correlate $\tilde{\kappa}_{WW}$ and  $\tilde{\kappa}_{ZZ}$ is not possible. 

Finally we also have the correlation,
\bea
g^h_{Zf}&=& 2 \delta g^Z_{f}+\frac{2g Y_f \ttw^2}{\ctw}\left(\kappa_{WW} - \kappa_{\gamma\gamma} -  \kappa_{Z\gamma}\frac{\ctw}{\stw}\right)\nn\\&-&
\left(2(\delta g^Z_{u_L}-\delta g^Z_{d_L})-\frac{\sqrt{2}}{\ctw}g^h_{WQ}     \right)(T_3 +Y_f \ttw^2),
\eea
which can also be translated to a correlation between the coupling $g^h_{Z\textbf{p}}$ in \eq{ghzp} and those in the right hand side above.

\paragraph{Connection to anomalous Triple Gauge Couplings}

The operators in Table~\ref{opers} also contribute to anomalous Triple Gauge Couplings (TGC) as follows,
\bea
\delta g^Z_{1}&=&\frac{1}{2 \stw^2}\frac{\delta m_Z^2}{m_Z^2}\\
\delta\kappa_{\gamma}&=&\frac{1}{\ttw} \frac{v^2}{\Lambda^2} c_{HWB}\,.
\eea
Using the above equation together with Eq.~\ref{wilson} and Eq.~\ref{kaa} we can obtain the following correlations between the and TGCs and the Higgs couplings to gauge bosons, 
\bea
g^h_{WQ}&=&{\sqrt{2}  \ctw}\left(\delta g^Z_{u_L}-\delta g^Z_{d_L} -g \ctw\delta g^Z_{1}  \right)\label{tgc1}\\
\kappa_{WW} &=&\delta\kappa_{\gamma}+ \kappa_{\gamma\gamma} +  \kappa_{Z\gamma}\frac{\ctw}{\stw} \label{tgc2}\,.
\eea
While $Wh$ production at high energies constrains $g^h_{WQ}$, the linear combination in the right hand side of \eq{tgc1} is precisely the EFT direction constrained by high energy $WZ$ production. This connection between $Wh$ and $WZ$ production is a consequence of the Goldstone boson equivalence theorem as explained in Ref.~\cite{Franceschini:2017xkh}. In Sec.~\ref{tgc} we will use the above relations to compare our bounds with TGC bounds obtained from double gauge boson production.

\begin{figure*}[!t]
\begin{center}
\includegraphics[scale=0.4]{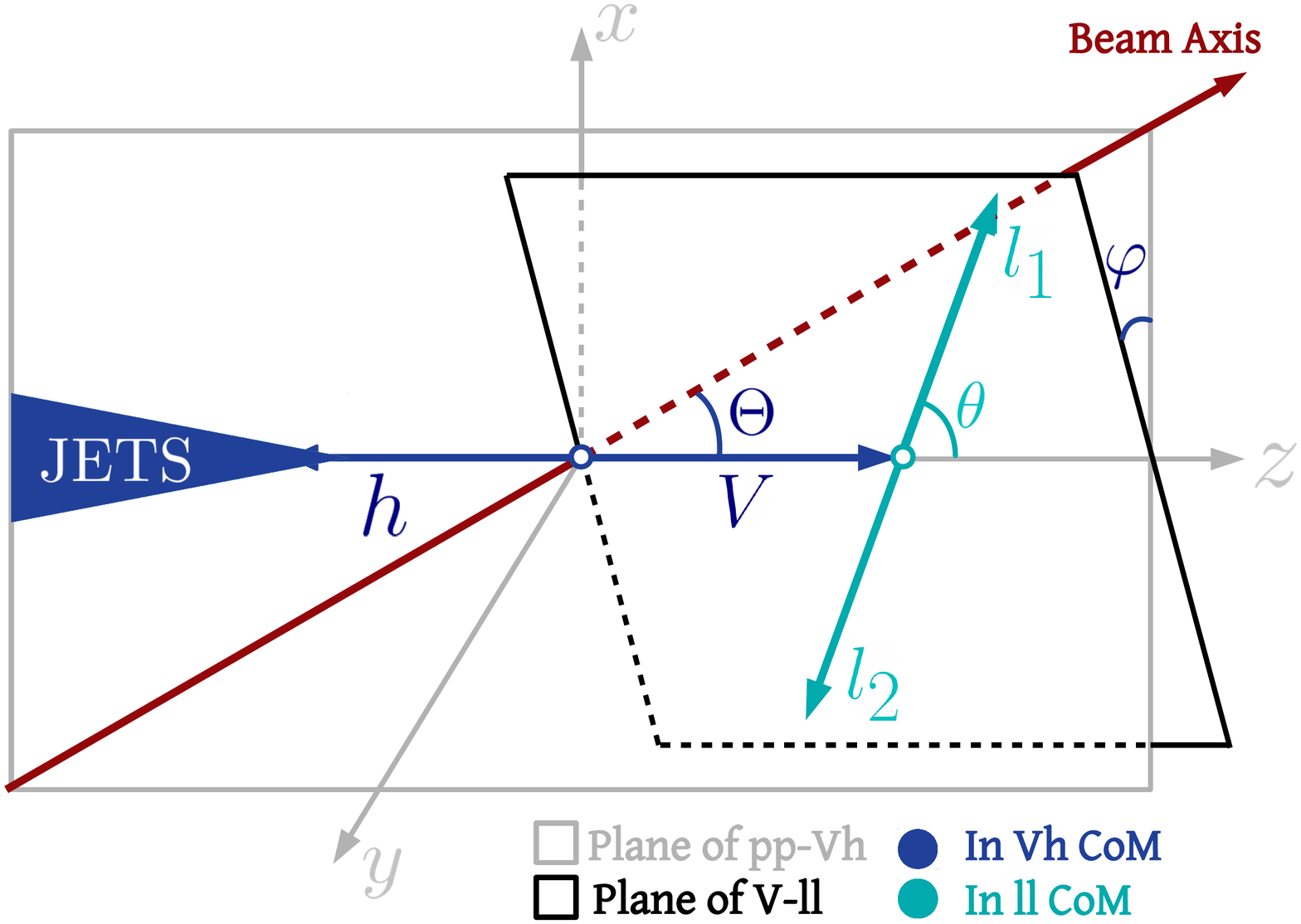}
\caption{Diagram showing the angles  that can completely characterise our final state. Note the use of  two different frames of reference: the CoM frame of the $Vh$ system (in which $\varphi$ and $\Theta$ are defined) and the CoM frame of $V$ (in which $\theta$ is defined). The Cartesian axes $\{x,y,z\}$ are defined by the $Vh$ centre-of-mass frame, with $z$ identified as the direction of the $V$-boson; $y$ identified as the normal to the plane of $V$ and the beam axis; $x$ defined so that it completes the right-handed set. }
\end{center}
\label{fig:angles}
\end{figure*}

\section{Angular moments for the   $pp \to V(\ell\ell)h(b\bar{b})$  process in the Dimension-6 SMEFT}
\label{angmom}

In this section we come to the central topic of this work and discuss how  the full angular distributions in the $pp \to V(\ell\ell)h(b\bar{b})$  processes, at a given energy, can be expressed in terms of a finite number of basis functions, both in the SM and D6 SMEFT. The corresponding coefficients of these functions are the so called angular moments for these processes. We start at the level of $ff \to V(\ell\ell)h(b\bar{b})$ and then discuss the experimental subtleties that arise in the extraction of these angular moments for $pp \to W^\pm(\ell\nu)h(b\bar{b})$  and $pp \to Z(\ell^+\ell^-)h(b\bar{b})$. As we will require the  two $b$-jets arising from the Higgs decay to form a fat jet in our analysis, we will effectively consider the three body final state of the fat jet and two leptons in this section.
\label{smeft}
\subsection{Angular moments at the $ff \to Vh$ level}
 The helicity amplitude formalism is the most convenient way to arrive at the full angular and energy  dependance of the   $ff \to V(\ell\ell)h(b\bar{b})$  amplitude. Starting at the 2$\to$2 level,  $f(\sigma)\bar{f}(-\sigma) \to Vh$, these helicity amplitudes are given by,
 
\bea
\label{amp}
\hspace{-5mm}&&{\cal M}_\sigma^{\lambda=\pm}= \sigma \frac{1+ \sigma \lambda \cos \Theta}{\sqrt{2}}G_V\frac{m_V}{\sqrt{\hat{s}}}\Bigg[1+\left(\frac{g^h_{Vf}}{g^V_f}+  \hat{\kappa}_{VV}- i \lambda \hat{\tilde{\kappa}}_{VV}\right)  \frac{\hat{s}}{2 m_V^2}  \Bigg]\nn\\
&&{\cal M}_\sigma^{\lambda=0}= -\frac{\sin \Theta}{2}G_V\Bigg[1+\delta \hat{g}^h_{VV}+  2\hat{\kappa}_{VV}+ \delta g^Z_f+\frac{{g}^h_{Vf}}{g^V_{f}} \left(-\frac{1}{2}+\frac{\hat{s}}{2 m_V^2}\right) \Bigg],
\eea
where,
\bea
\label{hats}
\hat{\kappa}_{WW}&=&\kappa_{WW},\nn\\
\hat{\kappa}_{ZZ} &=& \kappa_{ZZ}+\frac{Q_f e}{g^Z_f}\kappa_{Z\gamma},\nn\\
\hat{\tilde{\kappa}}_{ZZ} &=& \tilde{\kappa}_{ZZ}+\frac{Q_f e}{g^Z_f}\tilde{\kappa}_{Z\gamma},
\eea
and $G_{Z,W}=\frac{g g^Z_f}{\ctw}, \frac{g^2}{\sqrt{2}}$, $\lambda=\pm 1$ and $\sigma=\pm 1$ are, respectively, the helicities of the $Z$-boson and initial-state fermions, and $g^{Z}_{f}=g(T_3^f-Q_f \stw^2)/\ctw$ and $g^W_f=g/\sqrt{2}$; $\sqrt{\hat{s}}$  is the partonic centre-of-mass energy. The above expressions hold both for quark and leptonic initial states.  In \eq{amp} above, we have kept only the terms with leading powers of ${\sqrt{\hat{s}}}/m_V$ both for the SM and D6 SMEFT (the subdominant terms are smaller by, at least, factors of $m^2_V/{\hat{s}}$). We have, however, retained the next-to-leading EFT contribution for the $\lambda=0$ mode,  as an exception, in order to keep the leading effect amongst  the terms proportional to $\delta \hat{g}^h_{VV}$. The full expressions for the helicity amplitudes including the SMEFT corrections can be found  in Ref.~\cite{Nakamura:2017ihk}.  The above expressions assume that the fermion momentum is in the positive $z$-direction of the lab frame.  The expressions for the case where the anti-fermion has momentum in the positive $z$-direction can be obtained by making the replacement $\sigma \to -\sigma$. Above, we have not included the effect of a $Vff$ coupling deviation ($\delta g^V_f$ in \eq{anam}) above which we will incorporate at the end of this section.

It is worth emphasising that for both the SM and  D6 SMEFT,  only contributions up to the $J=1$ helicity amplitude appear. For the SM this is because the $ff \to Vh$ process is mediated by a spin-1 gauge boson. For the D6 SMEFT, in addition to diagrams with spin 1 exchange, there is also the contribution from the contact term in \eq{anam}. As this contact term is exactly the vertex that would arise by integrating out a heavy spin-1 particle, even in the D6 SMEFT only contributions up to $J=1$ exist. This fact will no longer be true at higher orders in the EFT expansion where higher-$J$ amplitudes will also get contributions.
 
At the $2 \to 3$ level, the amplitude can be most conveniently written in terms of $ {\varphi}$ and $ {\theta}$, the azimuthal and  polar angle of the  of the  negatively charged lepton for $V=W^-,Z$ and the neutrino for $V=W^+$, in the $V$ rest frame in the coordinate system defined in Fig.~\ref{fig:angles},
\bea
\label{helamp}
{\cal A}( \hat{s},  \Theta,  {\theta},  {\varphi})=\frac{-i g^V_\ell}{\Gamma_V}\sum_\lambda {\cal M}_\sigma^\lambda (\hat{s},\Theta)d^{J=1}_{\lambda,1}({\theta}) e^{i \lambda \hat{\varphi}},
\eea
where $g^V_\ell$ is defined below \eq{amp},  $\Gamma_V$ is the $V$-width, and $d^{J=1}_{\lambda,1}(\hat{\theta})$ are the Wigner functions,
\bea
\label{dfunc}
d^{J=1}_{\pm1,1}=\tau\frac{1\pm \tau \cos \theta}{\sqrt{2}},~d^{J=1}_{0,1}=\sin \theta,
\eea
with $\tau$ being the lepton helicity. We have assumed a SM  amplitude for the $V$-decay; modifications due to a $V\ell\ell$ coupling deviation will be included at the end of this section. For $V=W^\pm$ we always have $\tau=-1$. We can now obtain the squared amplitude with the full angular dependence using Eq.(\ref{amp}-\ref{dfunc}), 
\bea
\label{sumsquare}
\sum_\tau|{\cal A}(\hat{s},  \Theta, {\theta}, {\varphi})|^2=\sum_i a_{i}(\hat{s}) f_{i} (\Theta, {\theta}, {\varphi})\,,
\eea
where we have summed over the final lepton helicity.   The $f_{i} (\Theta, {\theta}, {\varphi})$ are the 9 functions we obtained by squaring  the sum of the  3 helicity amplitudes in the right-hand side of \eq{helamp}, see also~\cite{Collins:1977iv, Hagiwara:1984hi, Goncalves:2018ptp}. Explicitly these are,
\bea
    f_{LL} &=& S_\Theta^2 S_\theta^2,\nn\\
    f^1_{TT} &=& C_\Theta C_\theta,\nn\\
    f^2_{TT} &=& (1+C_\Theta^2)(1+ C_\theta^2),\nn\\
    f^1_{LT} &=& C_\varphi S_\Theta S_\theta,\nn\\
    f^2_{LT} &=& C_\varphi S_\Theta S_\theta C_\Theta C_\theta,\nn\\
    \tilde{f}^1_{LT} &=& S_\varphi S_\Theta S_\theta,\nn\\
    \tilde{f}^2_{LT} &=& S_\varphi S_\Theta S_\theta C_\Theta C_\theta,\nn\\
    f_{TT'} &=& C_{2\varphi}S_\Theta^2 S_\theta^2,\nn\\
    \tilde{f}_{TT'} &=&  S_{2\varphi}S_\Theta^2 S_\theta^2\,,
    \label{funcs}
\eea
where $S_\alpha= \sin \alpha,\, C_\alpha= \cos \alpha$. The subscripts of the above functions denote the $V$-polarisation of the two interfering amplitudes, with $TT'$ denoting the interference of two transverse amplitudes with opposite polarisations. The corresponding coefficients $a_i$ are the so-called angular moments for this process, which completely characterise the multidimensional angular dependance of this process at a given energy $\hat{s}$.  The expressions for these angular moments in terms of the vertex couplings in \eq{anam} are given in Table~\ref{coefs}. Note the factor,
\bea
\epsilon_{RL}=\frac{(g^V_{l_R})^2-(g^V_{l_L})^2}{(g^V_{l_R})^2+(g^V_{l_L})^2}\,,
\eea
in some of the moments, which arises from the sum over $\tau$ in \eq{sumsquare}.

It is worth emphasising an important conceptual point here. The cross-helicity moment functions, \textit{i.e.}, the last six functions in \eq{funcs}, integrate to zero over the full phase space of the $V$-decay products. This is expected as the two amplitudes corresponding to different helicities at the level of the $V$-boson cannot interfere. If we look at the phase space of the decay products differentially, however, the corresponding angular moments  carry very useful information. As one can verify from Table~\ref{coefs}, for instance, the leading contribution of the $\kappa_{ZZ}$ ($\tilde{\kappa}_{ZZ}$) coupling is to to the moment $a^2_{LT}$ ($\tilde{a}^2_{LT}$). As pointed out in Ref.~\cite{Banerjee:2019pks}, this effect can be recovered only if we study the triple differential with respect to all three angles, \textit{i.e.}, an integration over any of the three angles makes the basis functions $f^2_{LT}$ and $\tilde{f}^2_{LT}$ vanish. This is an example of an `interference resurrection' study, see also Refs.~\cite{Hagiwara:1986vm, azatov, panico, azatov2, Azatov:2019xxn}, where interference terms absent at the inclusive level are `recovered' by analysing the phase space of the decay products differentially.

It is possible that not all of these angular moments will be relevant or observable for a given initial and final state. Before considering in detail the case of the $pp \to V(ll)h$ process,  our main focus, let us briefly comment on which of these angular moments are accessible to lepton colliders. For the $e^+ e^- \to Z(\ell^+ \ell^-)h$ process in lepton colliders, all nine angular moments can be measured. However, three of them, namely $a^1_{TT}$, $a^1_{LT}$ and $\tilde{a}^1_{LT}$, are  suppressed by the factor  of $|\epsilon_{RL}|= 0.16$, which is  accidentally small due to the numerical closeness of the couplings $g^Z_{l_L}$ and $g^Z_{l_R}$.

Let us now compare our method, that parametrises the tree-level analytical amplitude in terms of angular moments,  to other methods   that construct observables/discriminants  using the full analytical amplitude such as the Matrix Element Likelihood Analysis (MELA) \cite{Anderson:2013afp} framework which is closely related to Optimal Observables~\cite{oo1,oo2,oo3} and  the Matrix-Element  Method~\cite{matrix, Gainer:2013iya}. These  approaches are similar in spirit to ours but  in all these cases the amplitude is expressed in terms of amplitude coefficients that are ultimately anomalous couplings or Wilson coefficients. This makes the corresponding observables more complicated and  less intuitive.  The optimal observable for a given coupling, for instance, will involve the full interference term due to that coupling and will be a linear combination over many moments. It will thus have a complicated distribution that cannot be easily visualised.  On the other hand, our approach using angular moments is very transparent physically. If moment shows  a deviation we can pinpoint the experimental distribution as well as the helicity amplitudes that are being affected. 


\begin{table}[t]
\small
\centering
\begin{tabular}{||c|c||} 
\hline
\rule{0pt}{4ex} $a_{LL}$& $\frac{{\cal G}_V^2}{4}\Big[1+2\delta \hat{g}^h_{VV}+ 4\hat{\kappa}_{VV}+ 2 \delta g^Z_f+\frac{{g}^h_{Vf}}{g^V_{f}}(-1+4 \gamma^2)\Big]$\\
\rule{0pt}{4ex} $a^1_{TT}$& $\frac{{\cal G}_V^2 \sigma \epsilon_{ RL}}{2\gamma^2}\Big[1+4\left(\frac{{g}^h_{Vf}}{g^V_{f}}+\hat{\kappa}_{VV}\right)\gamma^2\Big]$\\
\rule{0pt}{4ex} $a^2_{TT}$& $\frac{{\cal G}_V^2}{8\gamma^2}\Big[1+4\left(\frac{{g}^h_{Vf}}{g^V_{f}}+\hat{\kappa}_{VV}\right)\gamma^2\Big]$\\
\rule{0pt}{4ex} $a^1_{LT}$ & $-\frac{{\cal G}_V^2 \sigma\epsilon_{ RL}}{2\gamma}\Big[1+2\Big(\frac{2{g}^h_{Vf}}{g^V_{f}}+\hat{\kappa}_{VV}\Big){\gamma^2}\Big]$\\
\rule{0pt}{4ex} $a^2_{LT}$ & $-\frac{{\cal G}_V^2}{2\gamma}\Big[1+2\Big(\frac{2{g}^h_{Vf}}{g^V_{f}}+\hat{\kappa}_{VV}\Big){\gamma^2}\Big]$\\
\rule{0pt}{4ex} $\tilde{a}^1_{LT}$ & $-{\cal G}_V^2 \sigma\epsilon_{ RL} \hat{\tilde{\kappa}}_{VV} \gamma$\\
\rule{0pt}{4ex} $\tilde{a}^2_{LT}$ & $-{\cal G}_V^2  \hat{\tilde{\kappa}}_{VV}\gamma$\\
\rule{0pt}{4ex} $a_{TT'}$&$\frac{{\cal G}_V^2}{8\gamma^2}\Big[1+4\left(\frac{{g}^h_{Vf}}{g^V_{f}}+\hat{\kappa_{VV}}\right)\gamma^2\Big]$\\
\rule{0pt}{4ex} $\tilde{a}_{TT'}$  & $\frac{{\cal G}_V^2}{2}  \hat{\tilde{\kappa}}_{VV}$\\
\hline
 \end{tabular}
\caption{Expressions for the angular moments as a function of the different anomalous couplings in \eq{anam}  up to linear order. Contributions subdominant in $\gamma=\sqrt{\hat{s}}/(2 m_V)$ are neglected, with the exception of the next-to-leading EFT contribution to $a_{LL}$, which has been retained in order to keep the leading effect of the $\delta \hat{g}^h_{VV}$ term. The factor  $\epsilon_{ RL}$ is defined in text and ${\cal G}_V= g g^V_f \sqrt{(g^V_{l_L})^2+(g^V_{l_R})^2}/(\ctw \Gamma_V)$,  $ \Gamma_V$ being the $V$-width. The SM part of our results can also be found in~\cite{barger}.}
\label{coefs}
\end{table}

 \begin{figure}[!h]
\begin{center}
  \subfigure[]{\includegraphics[width=0.45\textwidth]{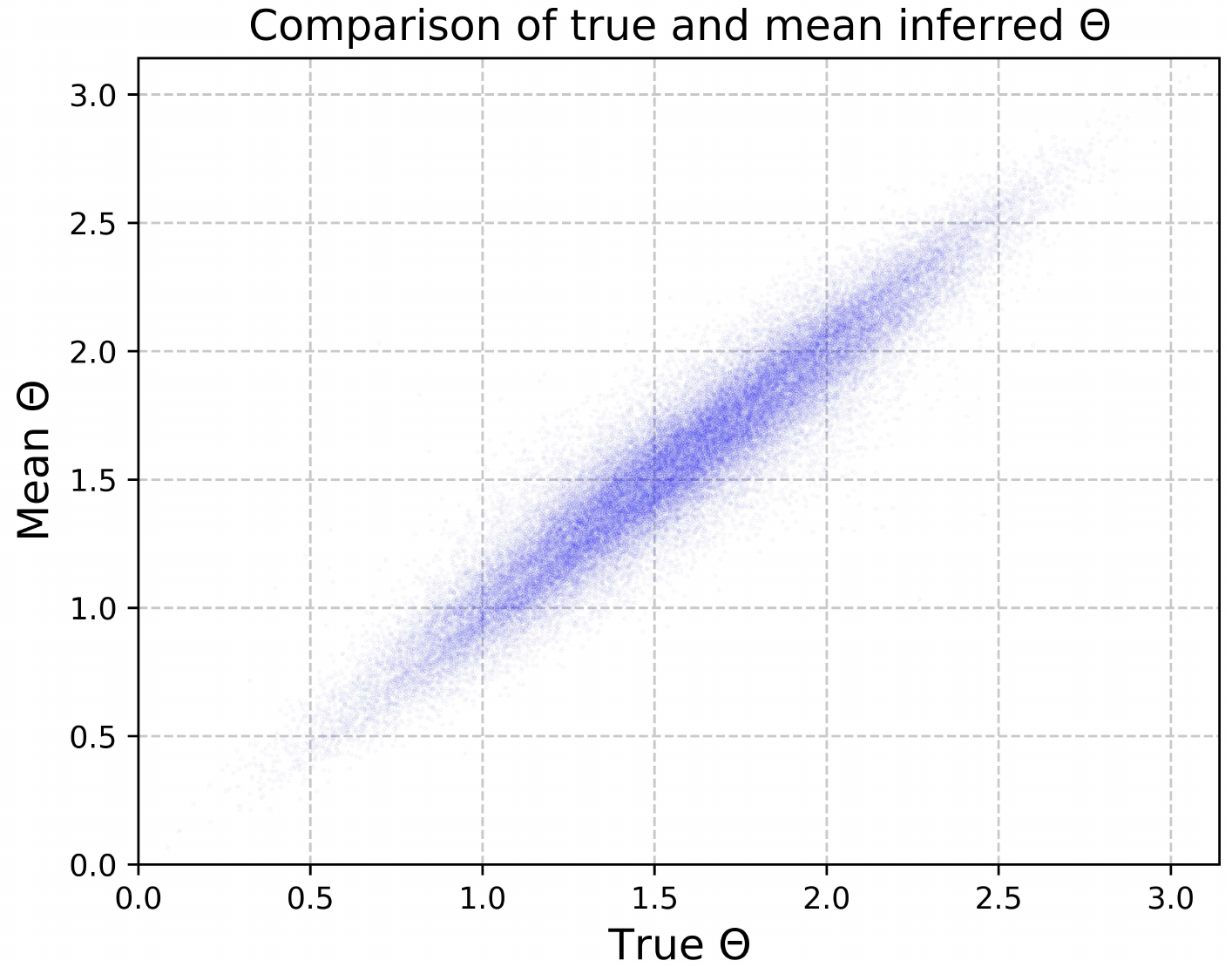} \label{BT}}\\
  \subfigure[]{\includegraphics[width=0.45\textwidth]{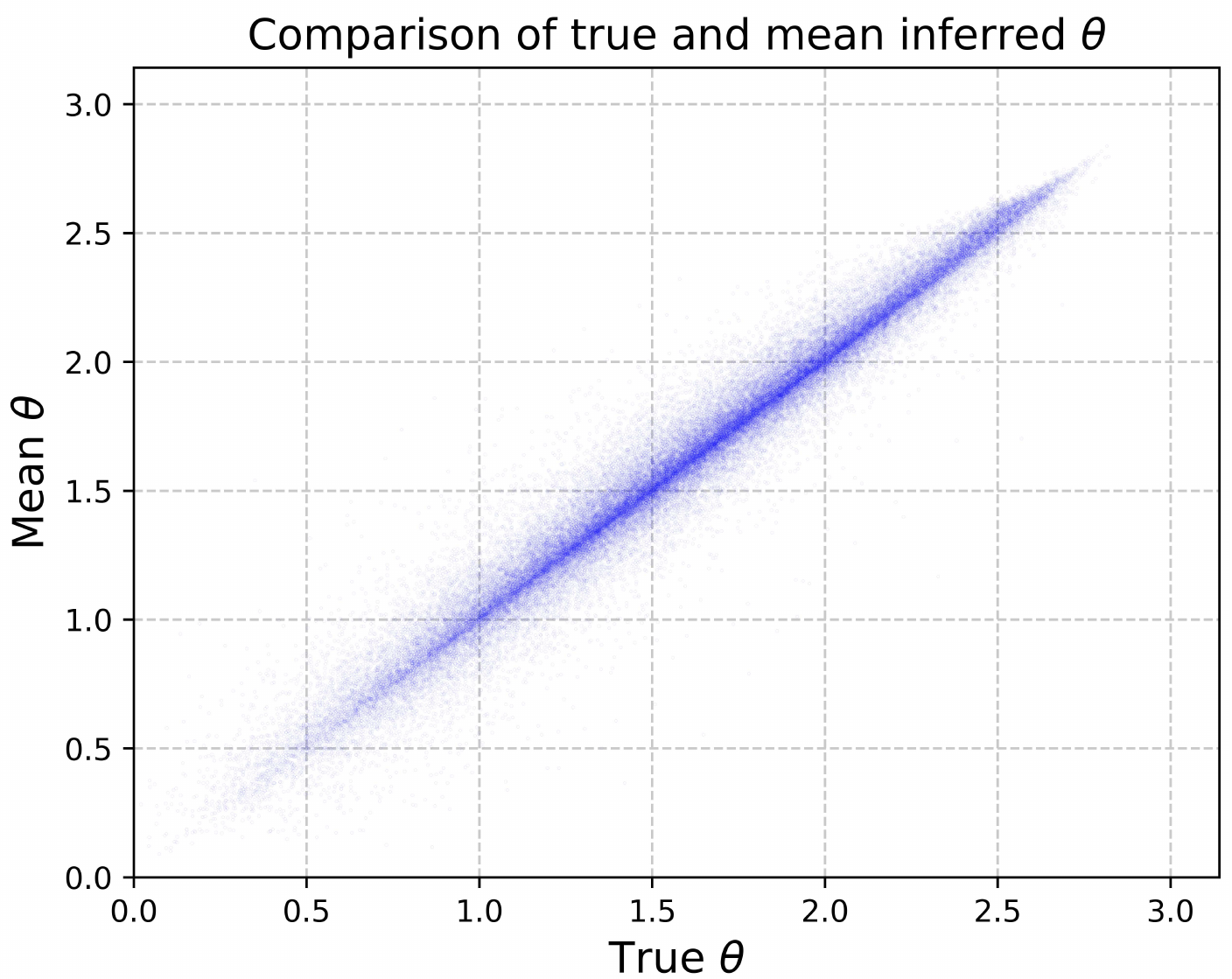} \label{ST}}\\
    \subfigure[]{\includegraphics[width=0.45\textwidth]{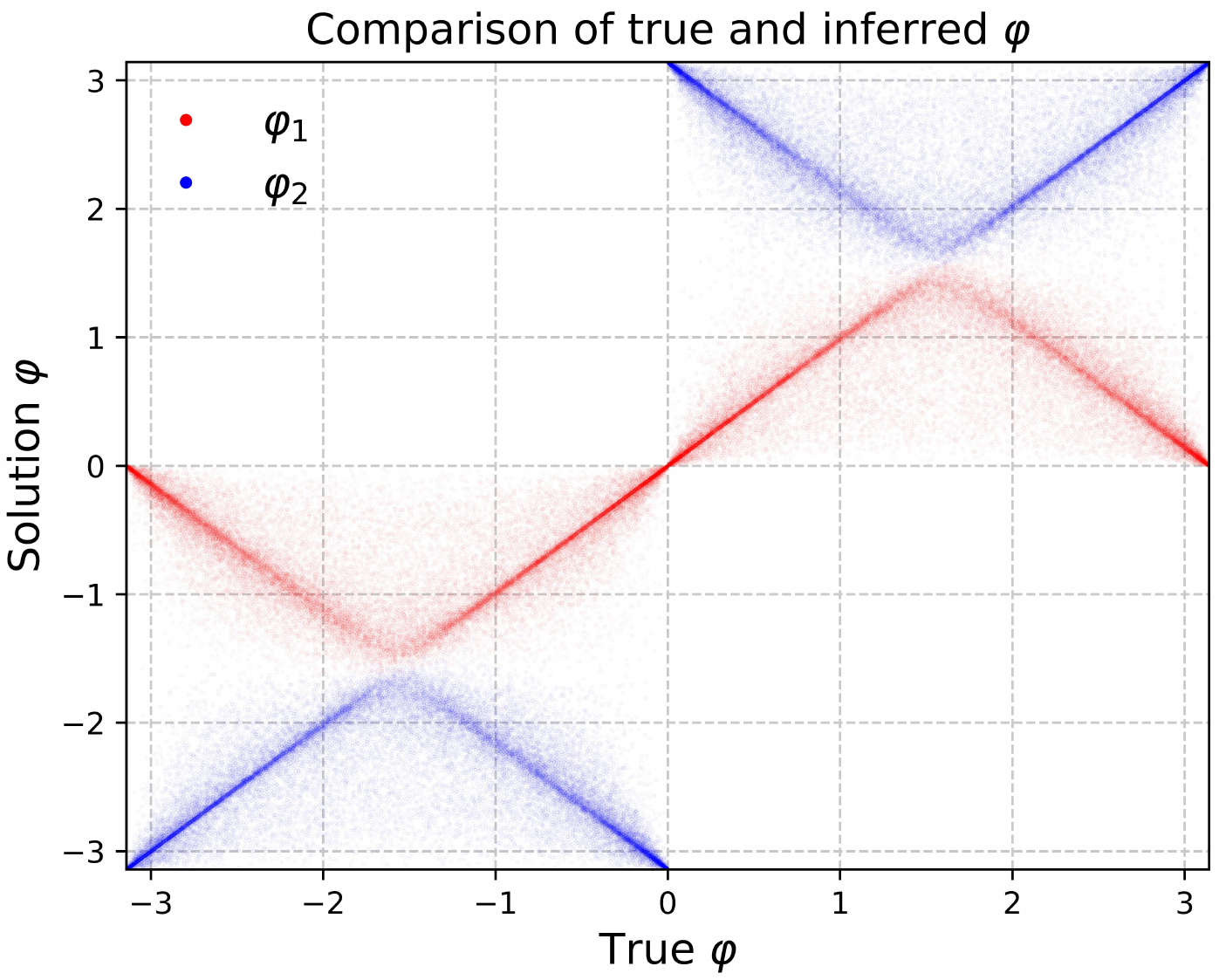} \label{PHI}}
 \caption{In the $W^\pm(l\nu)h(b\bar{b})$ process, the ambiguity in the $z$ momentum of the neutrino leads to two possible values of each of the three angles. Plots (a) and (b) above show the scatter plot for the mean of the solutions for  $\Theta$ and $\theta$ vs the true value.  Plot (c) includes two solutions for $\varphi$ (shown in red and blue) in a scatter plot vs the true value.}
 \end{center}
\end{figure}

\subsection{Angular moments for  the $pp \to Z(\ell\ell)h(b\bar{b})$ process}
\label{zh}

The first thing to note about the LHC  is that the direction of the quark is not always in the same direction in the lab frame. The expressions in Table~\ref{coefs} are for the case where the quark moves in the positive $z$-direction. For the other case where the antiquark momentum is in the $z$-direction, as stated below \eq{hats}, one can obtain the corresponding expressions for the angular moments by making the substitution $\sigma \to -\sigma$.  The angular moments $a^1_{TT}$, $a^1_{LT}$ and $\tilde{a}^1_{LT}$ thus vanish once we average over both these possibilities.  It should  be possible to recover some of this information by keeping track of the direction of the $Zh$ system as this is strongly correlated with the direction of the quark as the (valence) quark is generally more energetic than the anti-quark at high invariant masses. Thus, if the dataset is split into two parts according to the direction of the $Zh$ system it should be possible to extract these three moments also .  We will explore this possibility in future work.\footnote{We thank the anonymous referee for suggesting this idea.}

We are thus left with the 6 moments. At high energy, $a_{LL}$ dominates over all other moments in the SM. The largest BSM contribution at high energies is also to $a_{LL}$ from the linear combination $g^h_{Z\textbf{p}}$, see \eq{ghzp}, that arises from  averaging over the  initial state flavour and polarisation~\cite{Banerjee:2018bio}. The contribution due to  $g^h_{Z\textbf{p}}$ grows quadratically with energy and this coupling can thus be measured very precisely as we will see in Sec.~\ref{results}, this was also discussed in detail in   Ref.~\cite{Banerjee:2018bio}.

Once $g^h_{Z\textbf{p}}$ has been precisely measured we can use the remaining information in the angular moments to constrain the coupling $\delta \hat{g}^h_{ZZ}$ and the linear combinations,
\bea
\label{kappap}
\kappa^{\textbf{p}}_{ZZ}&=&\kappa_{ZZ}+0.3~\kappa_{Z\gamma}\nn\\
\tilde{\kappa}^{\textbf{p}}_{ZZ}&=&\tilde{\kappa}_{ZZ}+0.3~\tilde{\kappa}_{Z\gamma}\,,
\eea
that enter, respectively, the $CP$-even and odd  angular moments at the  $pp \to Z(\ell\ell)h(b\bar{b})$ level. The coefficient of $\kappa_{Z\gamma}$ and $\tilde{\kappa}_{Z\gamma}$ above arise again by appropriately averaging  \eq{hats} over the initial-state  flavours and polarisations.  Recall, however, that there is a very strong bound on $\kappa_{Z\gamma}$, see Sec.~\ref{eft}, so that the above linear combination effectively reduces to only $\kappa_{ZZ}$ to a very good approximation.

Consider now the angular moment  $a^2_{TT}$ and the  contribution to  $a_{LL}$ sub-dominant in $\gamma$, see Table~\ref{coefs}, which is unconstrained even after the strong bound on  $g^h_{Z\textbf{p}}$. First of all, the total rate of the  $pp \to Z(l^+l^-)h(b\bar{b})$ process depends only on the two moments  $a_{LL}$ and $a^2_{TT}$ as all other non-vanishing moments are coefficients of cross-helicity terms that vanish upon integration over $\varphi$, see \eq{funcs}. The rate itself can constrain a linear combination of $\delta \hat{g}^h_{ZZ}$ and   $\kappa^{\textbf{p}}_{ZZ}$. Additionally,   these two moments also carry the information of the joint  distribution of the events with respect to $(\theta, \Theta)$, which, along with the total rate, can  in principle be used to constrain  $\delta \hat{g}^h_{ZZ}$ and   $\kappa^{\textbf{p}}_{ZZ}$ simultaneously.  We find in our final analysis, however,  that the joint  $(\theta, \Theta)$ distribution in the events surviving our cuts is not very effective in simultaneously constraining these couplings. The main reason for this is that   the $\Theta$-distribution gets distorted with respect to the original theoretical form   because of the experimental cuts necessary for our boosted Higgs analysis. In particular, we require $p^h_T >150$ GeV, which eliminates  forward events. Another effect that could further distort the distribution is radiation of hard jets.\footnote{ If required, this effect can be mended by applying an active boost of the $ZH$ system to be on the collision axis, or by requiring that the transverse momentum of  all the final-state particles, excluding additional jets, is small compared to the hard scale of the event. The latter  is preferable compared to a jet veto as it avoids jet reconstruction uncertainties~\cite{Franceschini:2017xkh}.} As  $\theta$ and $\Theta$ appear in a correlated way in the amplitude, these effects also deform the $\theta$-distribution, but to a smaller extent. For this reason,  as discussed in Sec.~\ref{alternate}, we  will  isolate $a_{LL}$ and $a^2_{TT}$ using only the $\theta$-distribution in our final analysis,  in order to obtain better bounds.
 \begin{figure}[!t]
\begin{center}
  \subfigure[]{\includegraphics[width=0.55\textwidth]{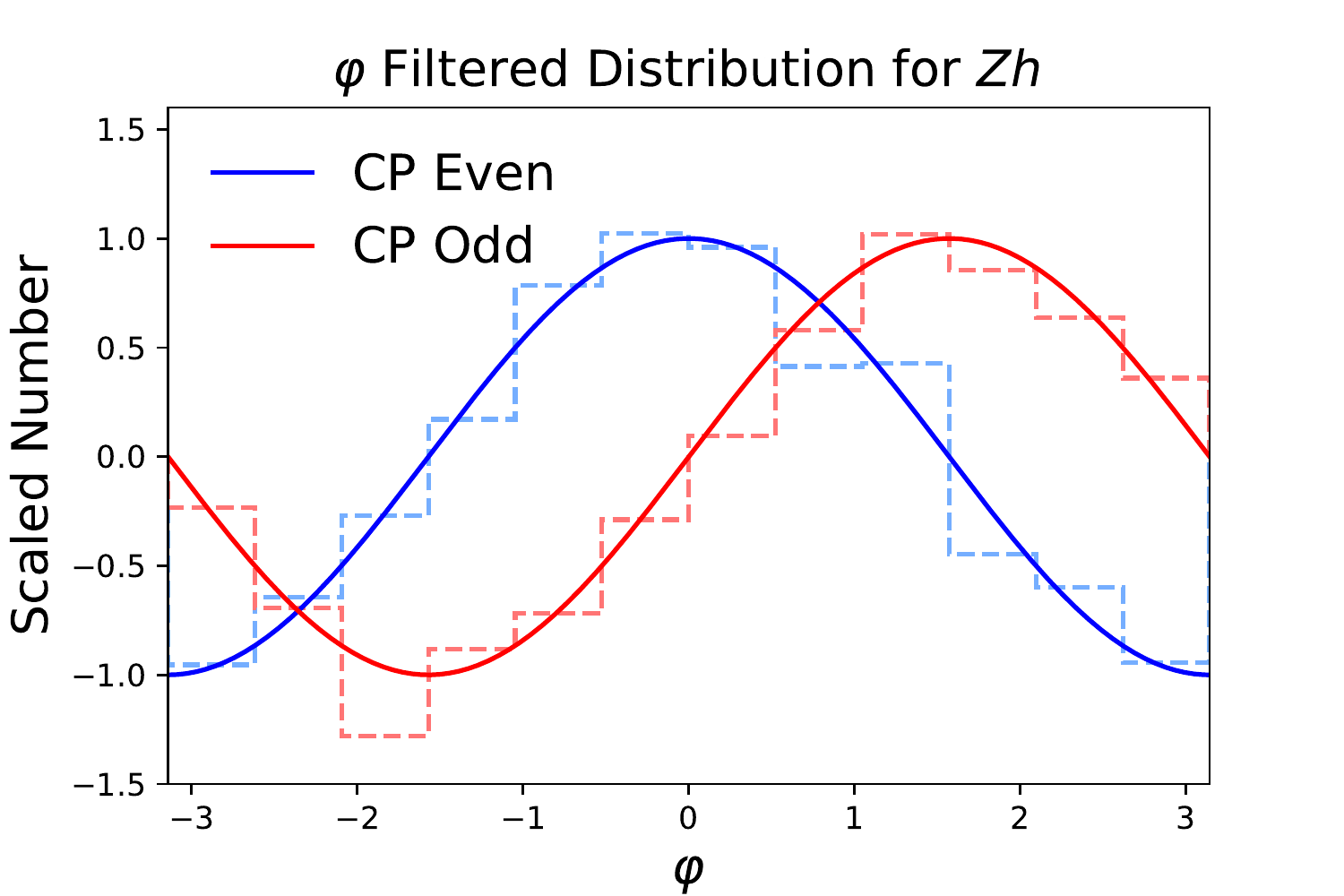} \label{zphi}}
  \subfigure[]{\includegraphics[width=0.55\textwidth]{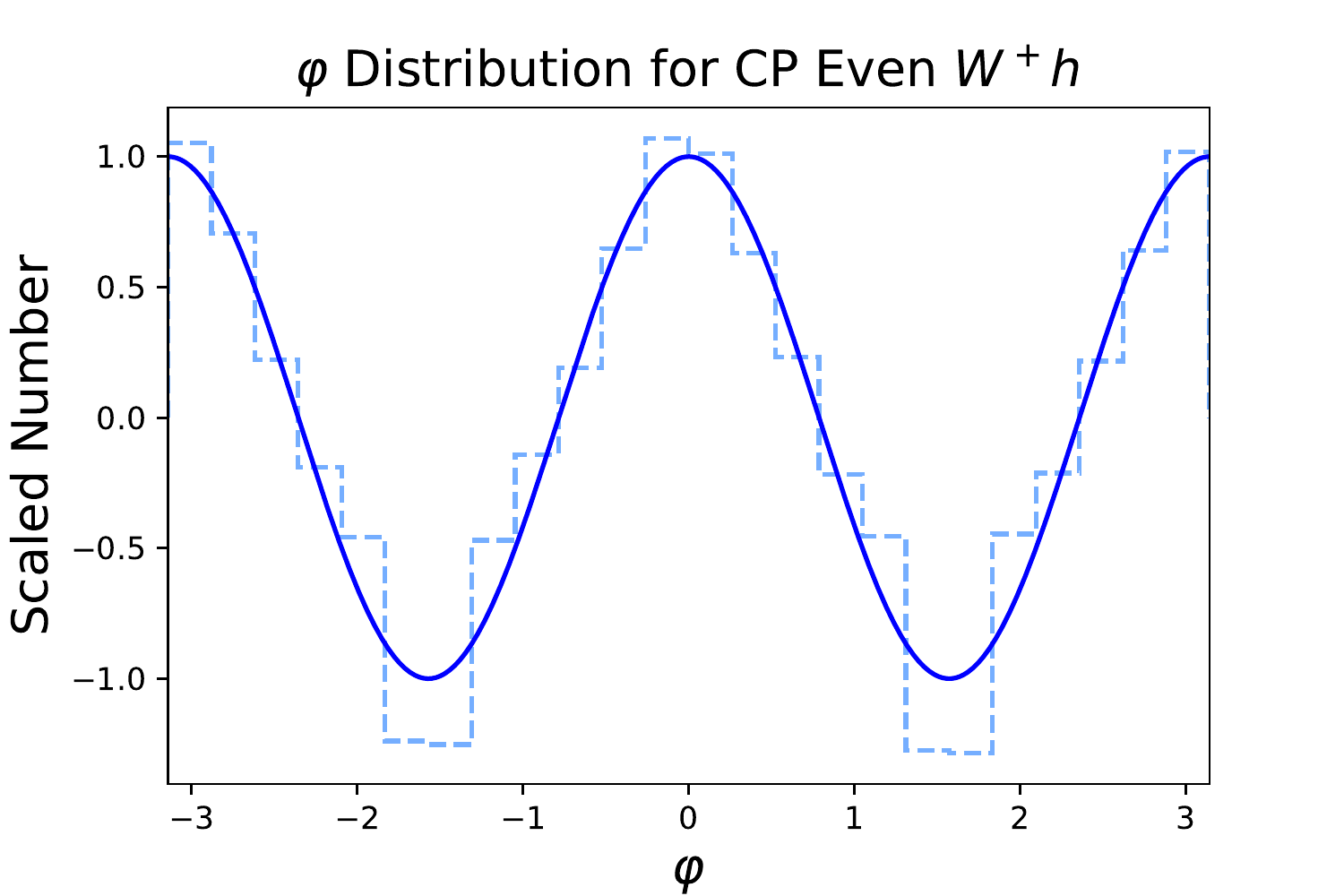} \label{wphi}}
    \caption{(a) Weighted $\varphi$-distributions for two different Monte-Carlo samples for the $Zh$ mode with the EFT couplings, $\kappa_{ZZ}$ and $\tilde{\kappa}_{ZZ}$, respectively,  turned on. The events used include showering and hadronisation and are those passing all selection cuts. To show the effect of the angular moments, $a^1_{LT}$ and $\tilde{a}^1_{LT}$, we take the weight of each event to be the sign of $\sin(2\theta)\sin(2\Theta)$. We then show the histogram with respect to  $\varphi$ and obtain the expected shapes for the two samples; (b) Regular $\varphi$-distributions for a Monte-Carlo sample for the $Wh$ mode with a non-zero value for the EFT coupling $\kappa_{WW}$. We see the effect of the angular moment $a_{TT'}$, the only angular moment that survives after integrating over $\theta$ and $\Theta$, and averaging over the two solutions. The events used are those passing all cuts.  The angular moment $\tilde{a}^1_{LT}$ can also be extracted in $Wh$ production but its effect can be seen only in a weighted distribution like in (a).}
 \end{center}
\end{figure}

 Much more reliable are the $\varphi$ distributions, which preserve their original shape to a large extent. We show in Fig.~\ref{zphi}, for instance, the  $\varphi$ distributions corresponding to an enhanced $a^2_{LT}$ and $\tilde{a}^2_{LT}$, for events that  include the effect of jet radiation and pass all experimental cuts to be described in Sec.~\ref{collider}.  We see the expected sinusoidal and cosinusoidal $\varphi$-dependances despite all these effects.
 
   The information for the $\varphi$-dependance is carried by  the angular moments $a^2_{LT}$ and $a_{TT'}$ in the $CP$-even case, which can be measured to   constrain the linear combination $\kappa^\textbf{p}_{ZZ}$, assuming  again that $g^h_{Vf}$ is already precisely constrained.  Among these, as identified in Ref.~\cite{Banerjee:2019pks}, the leading contribution is from $a^2_{LT}$, as it is larger relative to $a_{TT'}$ by a factor of $\gamma$, see Table~\ref{coefs}. This moment  provides the strongest bound on the above linear combination in our analysis but can be accessed only by looking at the joint distribution of $(\theta, \Theta, \varphi)$. A standard analysis that integrates over any of these three angles would miss this effect completely.

Finally the $CP$-odd coupling, $\tilde{\kappa}^{\textbf{p}}_{ZZ}$, cannot be constrained without using $\varphi$ information contained in $\tilde{a}^2_{LT}$ and $\tilde{a}_{TT'}$. Again, the leading effect contained in $\tilde{a}^2_{LT}$ is highly non-trivial and can only be accessed by utilising the triple differential distribution with respect to $(\theta, \Theta, \varphi)$.

Before moving to the next subsection, we would like to comment that the distortion of the distribution due to experimental cuts and jet radiation does not invalidate our analysis.    That is to say, while these effects perhaps reduce our sensitivity compared to the idealised case, as we will discuss later, these effects will  already be factored into our  uncertainty estimates. Moreover, our final analysis  does not depend too much on the precise shape of the $\Theta$-distribution, as we rely more on the $\theta$ and especially $\varphi$ distributions. 

 \subsection{Angular moments for  the $pp \to W(\ell\ell)h(b\bar{b})$ process}
 \label{wh}

Much of the discussion in the previous section is  also relevant here. Once again averaging over the initial quark-antiquark direction gets rid of the angular moments $a^1_{TT}$, $a^1_{LT}$ and $\tilde{a}^1_{LT}$.\footnote{As in the previous section keeping track of the direction of the $Wh$ system will help us recover some of this information. In this case, however, there is the further complication that this direction is unknown because of the two-fold ambiguity in the $p_z$ of the neutrino. Thus only  events. where both solutions for the neutrino $p_z$ yield the same $Wh$ direction. can be used to recover the effect of these moments.} The high energy amplitude is again dominated by $a_{LL}$ both in the SM and EFT. In the EFT case, the quadratically growing contribution due to $g^h_{WQ}$  can be used to  strongly constrain it. The discussion about   the distortion of the $\Theta$-distributions and its effect on extracting the moments $a_{LL}$ and $a^2_{TT}$ also holds  for this case.

The main difference from  $pp \to Z(\ell\ell)h(b\bar{b})$ arises in the  $\varphi$-distributions. A complication arises from the fact  that  the neutrino four momentum is experimentally inaccessible. Imposing energy and momentum condition and assuming an on-shell $W$-boson yields two possible solutions for the neutrino four momentum, \textit{i.e.}, two solutions for the $z$-component of the neutrino momentum in the lab frame, the $p_T$ being equal for both solutions.  While $\Theta, \theta$ and the final-state invariant mass converge for the two solutions, especially at high energies~\cite{panico}, the values of $\varphi$ for the two solutions do not converge, and in fact are related to each other as  $\varphi_2 = \pi - \varphi_1$ to a very good approximation. In our analysis we average over $\Theta, \theta$ and the final-state invariant mass, but keep both $\varphi$ solutions with equal weight. This has the consequence that the functions $\cos \varphi$ and $\sin 2 \varphi$ vanish when averaged over these two possibilities, resulting in the vanishing of the moments $a^1_{LT}$,  $a^2_{LT}$ and $\tilde{a}_{TT'}$, see \eq{funcs}.

In Fig.~\ref{BT}-\ref{PHI} we show, for the three angles, a scatter plot between the truth and reconstructed values obtained after our collider analysis described in Sec.~\ref{collider}. For $\Theta$ and $\theta$, we use for the reconstructed value the mean of the two solutions, whereas for $\varphi$, we populate the scatter plot with both solutions. It is clear from Fig.~\ref{PHI} that we have  $\varphi_1+\varphi_2= \pi$ to a very good approximation. While Fig.~\ref{BT}-\ref{PHI} show that the angles can be reconstructed quite well, the procedure is not exact, as we have assumed that $W$ is on-shell and did not properly take into account radiation of hard extra jets. In fact, for some rare events the virtuality of the $W$-boson is so high that no real solutions exist for the neutrino $p_z$, if we assume an on-shell $W$-boson; we neglect such events in our analysis.

In Fig.~\ref{wphi} we show the $\varphi$-distribution for EFT events that finally survive the collider analysis discussed in Sec.~\ref{collider}. We again see the expected $\cos(2\varphi)$ shape corresponding to $a_{TT'}$, which is the only moment that survives integration over the other two angles and the averaging over the two solutions (see also~\cite{Delaunay:2013npa}).  The difference in the true and reconstructed distributions at $\varphi =\pm \pi/2$ is related to fact that we discard events where the neutrino four momentum solutions are complex~\cite{panico}.

So far we have not considered the effect of $Vff$, $Vll $and $hbb$ coupling deviations due to D6 operators. All these coupling deviations are like $\delta \hat{g}^h_{VV}$ in that they simply rescale the SM amplitude and thus all SM distributions. Their effect can thus be incorporated by making the replacement in Table~\ref{coefs} and elsewhere,
\bea
\delta \hat{g}^h_{VV}\to \delta \hat{g}^h_{VV}+\delta \hat{g}^h_{bb}+\frac{2 \delta g^V_{f}}{g^V_f} +\frac{2 \delta g^V_{l}}{g^V_l}.
\label{rescale}
\eea
Of the above couplings,  while the $\delta g^V_{f,l}$  couplings are very precisely constrained to be close to zero,  the effect of $\delta \hat{g}^h_{bb}$ cannot be ignored.

\section {The Method of Moments}
\label{moments}
\subsection{Basic idea}
\label{basic}
As we have seen in Sec.~\ref{angmom}, the squared amplitudes for our processes can be decomposed into a set of angular structures, $f_i(\Theta,\theta,\varphi)$, whose contributions are parameterised by the associated coefficients, the so-called angular moments, $a_i$. We would like to extract these coefficients in a way that best takes advantage of all the available angular information. In principle, this can be done by a full likelihood fit, but here we use the method of moments~\cite{Dunietz:1990cj, james, Beaujean:2015xea}. This method has its advantages -- especially if the number of events is not too large~\cite{Beaujean:2015xea}. This method involves the use of an analog of Fourier analysis to extract the angular moments. Essentially, we look for weight functions, $w_i(\Theta,\theta,\varphi)$, that can uniquely extract the coefficients, $a_i$, i.e.,
\begin{align}
&&\int_0^\pi d\theta \int_0^\pi d\Theta \int_0^{2\pi} d\varphi \sum_i (a_i f_i) w_j\sin\theta \sin\Theta = a_j,\nonumber\\
&\Rightarrow& \int_0^\pi d\theta \int_0^\pi d\Theta \int_0^{2\pi} d\varphi f_iw_j\sin\theta \sin\Theta =\delta_{ij}.
\label{ortho}
\end{align}
Assuming that the weight functions are linear combinations of the original basis functions,
\begin{equation}
w_i = \lambda_{ij}f_j,
\end{equation}
we can use \eq{ortho} to show that the matrix $\lambda_{ij}=M_{ij}^{-1}$, where,
\begin{equation}
\label{matrixM}
M_{ij}= \int_0^\pi d\theta \int_0^\pi d\Theta \int_0^{2\pi} d\varphi f_if_j\sin\theta \sin\Theta.
\end{equation}
For the set of basis functions in \eq{funcs}, the resulting matrix is given by,
\begin{equation}
M = 
\left(
\begin{array}{ccccccccc}
\frac{512 \pi }{225} & 0 & \frac{128 \pi }{25} & 0 & 0 & 0 & 0 & 0 & 0 \\
0 & \frac{8 \pi }{9} & 0 & 0 & 0 & 0 & 0 & 0 & 0 \\
\frac{128 \pi }{25} & 0 & \frac{6272 \pi }{225} & 0 & 0 & 0 & 0 & 0 & 0 \\
0 & 0 & 0 & \frac{16 \pi }{9} & 0 & 0 & 0 & 0 & 0 \\
0 & 0 & 0 & 0 & \frac{16 \pi }{225} & 0 & 0 & 0 & 0 \\
0 & 0 & 0 & 0 & 0 & \frac{16 \pi }{9} & 0 & 0 & 0 \\
0 & 0 & 0 & 0 & 0 & 0 & \frac{16 \pi }{225} & 0 & 0 \\
0 & 0 & 0 & 0 & 0 & 0 & 0 & \frac{256 \pi }{225} & 0 \\
0 & 0 & 0 & 0 & 0 & 0 & 0 & 0 & \frac{256 \pi }{225} \\
\end{array}
\right)\,,
\end{equation}
where we have organised the basis functions in the order in which they appear in \eq{funcs}.

It is convenient to go to a basis such that $M_{ij}$ and thus its inverse $\lambda_{ij}$, are diagonal. This can be achieved by an orthogonal rotation,
\begin{align}
\hat{f}_1&=\cos \beta f_{LL}-\sin \beta f^2_{TT},\nonumber\\
\hat{f}_3&=\sin \beta f_{LL}+\cos \beta f^2_{TT},
\end{align}
by an angle, 
\begin{equation}
\tan\beta = -\dfrac{1}{2}(5+\sqrt{29}). 
\end{equation}
In the new fully-orthogonal basis, $\vec{\hat{f}}=\{\hat{f}_1,f^1_{TT},\hat{f}_3, f^1_{LT},f^2_{LT}, \tilde{f}^1_{LT},\tilde{f}^2_{LT},f_8,f_9\}$, the rotated matrix $M\to \hat{M}$ reads,
\begin{equation}
\hat{M}=\mathrm{diag}\left(
\frac{64 \pi }{225} \xi_+,
\frac{8 \pi }{9},
\frac{64 \pi }{225}\xi_-,
\frac{16 \pi }{9},
\frac{16 \pi }{225},
\frac{16 \pi }{9},
\frac{16 \pi }{225},
\frac{256 \pi }{225},
\frac{256 \pi }{225}
\right)
\end{equation}
with $\xi_\pm = (53\pm9\sqrt{29})$. This is the matrix $\hat{\lambda}_{ij}^{-1}$, so that the weight functions in the rotated basis are,
\begin{equation}
\label{wi}
w_i = \hat{M}^{-1}_{ij}f_j.
\end{equation}
We are now able to convolute our event distributions with these weight functions to extract values for the coefficients in the new basis, 
\bea
\label{basis1}
\{\hat{a}_{1},a^1_{TT},\hat{a}_{3},a^1_{LT}, a^2_{LT}, \tilde{a}^1_{LT},\tilde{a}^2_{LT}, a_{TT'}, \tilde{a}_{TT'}\}
\eea
which can then be rotated back if we are interested in the moments in the original basis.

\subsection{Alternative weight functions for \texorpdfstring{$a_{LL}$}{aLL} and \texorpdfstring{$a^2_{TT}$}{a2TT} }
\label{alternate}

The above algorithm to extract the moments, systematically generates the set of weight functions, but this set is not unique. For instance, a function proportional to $\cos 2\varphi$ can also be the weight function for $f_{TT'}$. As we mentioned in Sec.~\ref{angmom}, the $\Theta$ distribution suffers distortions to its original shape due to experimental cuts and other effects. For this reason, the extraction of $a_{LL}$ and $a^2_{TT}$ using the weight functions derived above does not give optimal results. To avoid this, we can use weight functions only involving $\theta$ to extract these two moments. 

Let us integrate \eq{sumsquare} over the $\Theta$ and $\varphi$ to keep only the $\theta$ dependance,
\begin{align}
\int d\varphi d\Theta \sin \Theta \sum_\tau |{\cal A}(\hat{s}, \Theta, {\theta}, {\varphi})|^2&=a'_{LL} f'_{LL} ( {\theta})+a^{2'}_{TT} f^{2'}_{TT}({\theta})\nonumber\\
&=a'_{LL} \sin^2 \theta +a^{2'}_{TT} (1+\cos^2 \theta),
\end{align}
where $a'_{LL}$ and $a^{2'}_{TT}$ are related to the original moments $a_{LL}$ and $a^2_{TT}$ as follows,
\begin{align}
\label{conversion}
a'_{LL}= \frac{8\pi}{3}a_{LL},~~~~~~a^{2'}_{TT}= \frac{16\pi}{3}a^2_{TT}.
\end{align}
Now, following the steps in Sec.~\ref{basic}, we carry out a rotation,
\begin{align}
\hat{f'}_1&=\cos \beta' f'_{LL}-\sin \beta' f^{2'}_{TT},\nonumber\\
\hat{f'}_3&=\sin \beta' f'_{LL}+\cos \beta' f^{2'}_{TT},
\end{align}
to diagonalise the matrix in Sec.~\ref{matrixM}. In this case, the angle of rotation is given by $\tan \beta'=1$. In this basis, the weight functions are proportional to $\hat{f'}_1$ and $\hat{f'}_3$, given by,
\begin{align}
\hat{w}'_1(\theta)&=\hat{f}'_1(\theta)\frac{3(\sqrt{61}-9)}{16},\nonumber\\
\hat{w}'_3(\theta)&=\hat{f}'_3(\theta)\frac{3(\sqrt{61}+9)}{16}.
\end{align}
Convoluting the observed distribution with these weight functions yields $\hat{a}'_{1}$ and $\hat{a}'_{3}$, which can be rotated back to give ${a}'_{LL}$ and $\hat{a}^{2'}_{TT}$ and finally ${a}_{LL}$ and $\hat{a}^{2}_{TT}$ using \eq{conversion}. Using these alternative weight functions is equivalent to using only the information in the $\theta$-distribution to extract these two moments and ignoring the distorted $\Theta$ distribution. This will improve the final bounds we obtain in Sec.~\ref{results}. For clarity, we denote the full set of   angular moments in this particular orthonormal basis,
\begin{equation}
\label{basis2}
\{\hat{a}'_{1},a^1_{TT},\hat{a}'_{3},a^1_{LT}, a^2_{LT}, \tilde{a}^1_{LT},\tilde{a}^2_{LT}, a_{TT'}, \tilde{a}_{TT'}\}.
\end{equation}

Note that the other moment functions corresponding to  $a^2_{LT}, \tilde{a}^2_{LT}, a_{TT'}$ and $\tilde{a}_{TT'}$ also depend on $\Theta$ but we did not choose alternate weight functions for them because we checked that these moments can be accurately  determined despite the deformations in the $\Theta$-distributions. The reason for this is probably the fact that the $\varphi$-distributions are well preserved even after showering, hadronisation and experimental cuts and the moment functions include simple trigonometric  functions of $\varphi$, such as $\sin 2 \varphi$ and $\cos 2 \varphi$, that can be neatly separated just using the $\varphi$ distributions.As far as $a^2_{LT}$ and  $\tilde{a}^2_{LT}$ are concerned it  is impossible to chose  weight functions independent of $\theta$ and $\Theta$ because the corresponding functions vanish when integrated over these angles. It is still possible to accurately determine these angular moments because despite the deformations, the final  $\theta$ and $\Theta$ distributions are still odd under the two transformations $\Theta\to \pi-\Theta$ and $\theta \to \pi-\theta$ so that these angular moments can still be extracted by  convoluting the observed distributions with the existing weight functions.

\subsection{Extraction of angular moments and uncertainty estimate}
To obtain our sensitivity estimates for the anomalous couplings, we will take as the SM as the null-hypothesis and the  expected value of the angular moments would be given by $a^{SM}_i$.  We assume that the experiments would finally measure angular moments that deviate from the SM predictions because of the presence of the EFT couplings.  We are  interested in the deviation,  $(a^{EFT}_i-a^{SM}_i)$, where $a^{EFT}_i$ are the experimentally measured angular moments,
\begin{align}
\label{angsum}
a^{EFT}_i (M)= \sum_{n=1}^{\hat{N}} w_i(\Theta_n,\theta_n,\varphi_n)\,,
\end{align}
that are functions of the EFT couplings. Notice that the convolution in \eq{ortho} becomes a simple summation over all $\hat{N}$ events in the experimental dataset.

 In the absence of the true experimental dataset we will use  our simulated SM and EFT samples (see Sec.~\ref{collider}) to estimate  the expected value  of  $a^{EFT}_i$, $a^{SM}_i$ and the associated   statistical uncertainties. We will also take a flat systematic uncertainty on the SM prediction given by $\kappa_{\rm syst} a^{SM}_i$ where we will take $\kappa_{\rm syst}=0.05$ in this work. Again,  \eq{ortho} becomes a simple summation over all the events in our sample,
\begin{align}
\label{angsum}
a_i (M)= \frac{\hat{N}}{N}\sum_{n=1}^{N} w_i(\Theta_n,\theta_n,\varphi_n)\,,
\end{align}
where  depending on the case at hand we will use either the basis in \eq{basis1} or the one in \eq{basis2}  for our final analysis. In order to also take energy dependance into account, we have split the events into bins of the final-state invariant mass, with $M$ being the central value of a given bin. Here, $N=N(M)$ is the number of Monte-Carlo events in the sample and $\hat{N}=\hat{N}(M)$ the actual number of events expected, both in the particular invariant-mass bin for a given integrated luminosity.  Note that we have changed the normalisation of  the angular moments in \eq{angsum}; now $\sum_i {a}_i {f}_{i}$, in any given basis, yields the distribution of the actual number of events expected at a certain integrated luminosity and not the squared amplitude as in \eq{sumsquare}. For a sufficiently-large number of events, $N$, the weight functions, $w_i$, converge to a multivariate Gaussian distribution with a mean and covariance matrix given by,
\begin{align}
\bar{w_i}&=\frac{1}{N}\sum_{n=1}^N w_i(\Theta_n,\theta_n,\varphi_n)\,,\nonumber\\
\sigma_{ij}&=\frac{1}{N-1}\sum_{n=1}^N\left[w_i-\bar{w_i}\right] \left[{w_j}-\bar{w_j}\right].
\end{align}
We find that if we keep increasing $N$, as soon as it is large enough (say 100), the $\bar{w_i}$ and $\sigma_{ij}$ approach fixed values. In the orthonormal bases of \eq{basis1} and \eq{basis2} we find a covariance matrix that is nearly diagonal.

For a fixed $\hat{N}$, if we assume a diagonal covariance matrix, the angular moments in the orthonormal basis converge to Gaussians with mean and standard deviation given by,
\begin{align}
a_i\pm\delta a_i=\hat{N} \bar{w_i}\pm \sqrt{\hat{N} \sigma_{ii}}\,.
\end{align}
As a cross-check, we also computed the second term above, $\delta a_i$, by splitting our Monte-Carlo sample into parts with $\hat{N}$ events each and computing $a_i$ in each case; the standard deviation of the $a_i$ obtained matches the second term above very closely. This way of estimating the error also shows that any deformation of the original angular distribution due to experimental or QCD effects (see \eq{zh}), has been already factored into our uncertainty estimate.

To estimate the final uncertainty on the $a_i$ one must also consider the fact that, $\hat{N}$, the expected number of events in the given bin, itself fluctuates statistically. Finally there is the systematic uncertainty on the SM prediction that we mentioned above. Adding all these errors in quadrature we obtain, for the uncertainty in the SM Prediction, $a^{SM}_i$,
\begin{align}
\label{angerr}
\Sigma_i= \sqrt{\left(\left(\frac{\sqrt{\hat{N}}}{\hat{N}}\right)^2 +\kappa_{\rm syst}^2\right)(a^{SM}_i)^2+\hat{N} \sigma^{SM}_{ii}}.
\end{align}

\section{Collider Simulation}
\label{collider}
In this study, we take into account NLO QCD effects. We work under the \textsc{MG5\_aMC@NLO} \cite{Alwall:2014hca} environment to generate NLO events showered using \textsc{Pythia8}~\cite{Sjostrand:2001yu, Sjostrand:2014zea}. Inside this framework, real emission corrections are performed following the FKS subtraction method \cite{Frixione:1995ms}, whereas virtual corrections are done using the OPP reduction technique~\cite{Ossola:2006us}. The MC@NLO formalism~\cite{Frixione:2002ik} takes care of the matching between the LO matrix element and parton shower, thus avoiding double counting. Decay of heavy bosons has been carried out with the help of \textsc{MadSpin}~\cite{Artoisenet:2012st}, which retains spin information at tree-level accuracy. We construct our NLO model using \textsc{FeynRules}~\cite{Alloul:2013bka} and then employ \textsc{NLOCT}~\cite{Degrande:2014vpa} to compute the $UV$ and $R_2$ counterterms, which are required for the one-loop calculation. $UV$ counterterms are essential to remove ultraviolet divergences that appear at the loop level, whereas $R_2$ terms originate from the one-loop integrands that carry $(n-4)$-dimensional pieces in the numerators and $n$-dimensional terms in the denominators. As and when required, we manually insert the $R_2$ terms in the NLO model as the usage of publicly-available NLOCT version is restricted to renormalisable interactions only.

In this work, we focus on three different processes, \textit{i.e.}, $p p \to Zh$ and $p p \to W^{\pm} h$, with the Higgs decaying to a pair of $b$-quarks and the $Z/W$ decaying leptonically. Thus, for the $Zh \; (Wh)$ process, we study the $\ell^+\ell^- b \bar{b} \; (\ell \nu b\bar{b})$ final states, where $\ell = e, \mu, \tau$. The $q \bar{q} \to Z h$ and $q \bar{q}' \to W^{\pm} h$ processes are generated at NLO QCD, whereas the $g g \to Z h$ channel is generated at LO (which is at one loop). The following analyses are performed at 14 TeV centre-of-mass energy and the predictions are shown for the HL-LHC for an integrated luminosity of 3 ab$^{-1}$.

\subsection{The $Zh$ channel}

First we outline the generations of the signal and background samples for the $p p \to Zh \to b \bar{b} \ell^+ \ell^-$ analysis. While generating the signal samples, \textit{i.e.}, $q \bar{q} \to Z h$, we use the aforementioned NLO model file and interface it with {\sc Pythia8}. We choose dynamic renormalisation and factorisation scales, $\mu_F = \mu_R = m_{Zh}$. We choose NNPDF2.3@NLO as our parton distribution function (PDF) for the NLO signal samples. As mentioned above, for the NLO signal samples we use {\sc MadSpin}~\cite{Artoisenet:2012st} to decay the heavy bosons. This step is done at LO and hence we correct for the branching ratios following the Higgs working group recommendations. We follow Refs.~\cite{Banerjee:2018bio, Banerjee:2019pks} while generating the background samples. All background samples are generated at LO with NNPDF2.3@LO as the PDF. The dominant backgrounds comprises the $Zb\bar{b}$ and the irreducible SM $Zh$ production. For the $Zb\bar{b}$ production, we consider the tree-level mode as well as the $g g \to ZZ$ mode at one-loop. Furthermore, we consider reducible backgrounds like $Z+$ jets and the light jets are misidentified as $b$-tagged jets ($c$-jet misidentification is not considered separately), and the fully leptonic decay of $t\bar{t}$. Rather than performing a standard resolved analysis, where one would consider two separate narrow $b$-tagged jets, here we require a fat jet with its jet parameter $R=1.2$. We utilise a modified version of the BDRS algorithm~\cite{Butterworth:2008iy} in order to maximise sensitivity. This procedure helps us in maximising the signal by retaining extra radiations and in discriminating electroweak-scale resonant signals from strong QCD backgrounds, see also \cite{Soper:2010xk,Soper:2011cr}.

To briefly review the BDRS approach, the jets are recombined upon using the Cambridge-Aachen (CA) algorithm~\cite{Dokshitzer:1997in, Wobisch:1998wt} with a considerably large cone radius in order to contain the maximum number of decay products ensuing from a resonance. The jet clustering process is then read through backwards and one stops when the mass of a subjet, $m_{j_1} < \mu m_j$ with $\mu=0.66$, where $m_j$ is the mass of the fatjet. This step is called the {\textit{mass drop}} and is required to occur without a significant asymmetric splitting,
$$\frac{\text{min}(p_{T,j_1}^2,p_{T,j_2}^2)}{m_j^2}\Delta R_{j_1,j_2}^2>y_{\text{cut}},$$
where $y_{\text{cut}} = 0.09$. When this condition is not satisfied, the softer subjet, $j_2$, is removed from the list and the subjets of $j_1$ are subjected to the aforementioned criteria. This procedure is repeated iteratively until the aforementioned condition is met. This algorithm terminates when one obtains two subjets, $j_{1,2}$ which abide by the mass drop condition. However, the mass drop algorithm does not improve the resonance reconstruction significantly and more fine-tuning is necessary to segregate the signal from the background. A further step is performed: \textit{filtering}. In this algorithm, the constituents of the subjets $j_1$ and $j_2$ are further recombined using the CA algorithm but with a cone radius $R_{\text{filt}} = \text{min}(0.3, R_{b\bar{b}}/2)$. This algorithm chooses only the hardest three filtered subjets in order to reconstruct the resonance. In the original paper~\cite{Butterworth:2008iy}, the resonance in question is the SM-like Higgs boson and thus the hardest two filtered subjets are required to be $b$-tagged. In the present work, we find that the filtered cone radius $R_{\text{filt}} = \text{max}(0.2, R_{b\bar{b}}/2)$ performs better in reducing the backgrounds. As shown in Ref.~\cite{Butterworth:2008iy}, the filtering step significantly reduces the active area of the initial fatjet. Finally, we require the hardest two filtered subjets to be $b$-tagged with tagging efficiencies of 70\%. Moreover, the misidentification rate of light subjets faking as $b$-subjets is taken as 2\%.

One of our goals is to look for new physics effects in high-energy bins and hence it is imperative to generate the signal and background samples with certain generation-level cuts in order to improve statistics. For the $q \bar{q} \to Z h$ samples generated at NLO, we require a cut on the $p_T$ of the Higgs boson, $p_{T,h} > 150$ GeV. The $Zb\bar{b}$ and $t\bar{t}$ samples are generated with the following cuts: $p_{T,(j,b)} > 15$ GeV, $p_{T,\ell} > 5$ GeV, $|y_j| < 4$, $|y_{b/\ell}| < 3$, $\Delta R_{b\bar{b}/bj/b\ell} > 0.2$, $\Delta R_{\ell\ell} > 0.15$, $70 \; \text{GeV} < m_{\ell \ell} < 110$ GeV, $75 \; \text{GeV} < m_{b\bar{b}} < 155$ GeV and $p_{T,\ell\ell} > 150$ GeV. The $Zb\bar{b}$ sample is generated upon merging with an additional matrix element (ME) parton upon using the MLM merging scheme~\cite{Mangano:2006rw}. 
For the $Z+$ jets samples, we do not impose any invariant mass cuts in the jets. Furthermore, the sample is merged with three additional partons. Since the backgrounds are generated at LO, we use flat $K$-factors to bring them at a similar footing to the signal. For the tree-level $Zb\bar{b}$, one loop $g g \to ZZ$, one loop $ g g \to Z h$ and $Z+$ jets, we respectively use $K$-factor values of 1.4 (computed within \textsc{MG5$\_$aMC@NLO}), 1.8~\cite{Alioli:2016xab}, 2~\cite{Altenkamp:2012sx} and 1.13, computed within MCFM~\cite{Campbell:1999ah, Campbell:2011bn, Campbell:2015qma}.

A cut-based analysis has been done in Ref.~\cite{Banerjee:2018bio} and it has been shown that the prowess of a multivariate analysis exceeds that of a simple cut-and-count analysis. Thus, in this work we do not revisit the cut-and-count analysis and delve directly into the multivariate formulation. We start by constructing fatjets with cone radii of $R=1.2$. Furthermore, we require these fatjets to have $p_T > 80$ GeV and to lie within a rapidity, $|y| < 2.5$. We employ \textsc{FastJet}~\cite{Cacciari:2011ma} in constructing the jets. Moreover, we isolate the leptons ($e,\mu$) upon demanding that the total hadronic activity deposited around a cone radius of $R=0.3$ can at most be 10\% of its transverse momentum. The leptons are also required to have $p_T > 20 $ GeV and have rapidity, $|y| < 2.5$. In our setup, every non-isolated object is considered to be part of the fatjet construction. Before performing the multivariate analysis, we require each event to have exactly two oppositely charged same flavour (OSSF) isolated leptons. Moreover, we apply loose cuts on certain kinematic variables. We require the invariant mass of the leptons to be in the range $70 \; \text{GeV} < m_{\ell\ell} < 110$ GeV, the transverse momentum of the di-lepton system, $p_{T,\ell\ell} > 160$ GeV. We also require $\Delta R_{\ell\ell} > 0.2$~\footnote{$\Delta R = \sqrt{(\Delta \phi)^2 + (\Delta y)^2}$, where $\Delta \phi$ and $\Delta y$ are respectively the separation in azimuthal angle and rapidities of the two objects.}, $p_{T,\text{fatjet}} > 60$ GeV, the reconstructed Higgs mass, $95 \; \text{GeV} < m_h < 155$ GeV, $\Delta R_{b_i,\ell_j} > 0.4$ ($i=1,2$) and $\slashed{E}_T < 30$ GeV. We also require that there is at least one fat jet with at least two $B$-meson tracks, there are exactly two mass-drop subjets and at least three filtered subjets. We also require that the hardest two filtered subjets are $b$-tagged. Owing to the smallness of the $Z+$ jets and $t\bar{t}$ backgrounds compared to $Zb\bar{b}$, we train our boosted decision tree (BDT) upon only considering the NLO $Zh$ and the tree-level $Zb\bar{b}$ samples. We use the following variables to train the BDT: $p_T$ of both isolated leptons, $\Delta R$ between the $b$-subjets and the isolated leptons (four combinations), between the isolated leptons and also between the two $b$-subjets in the fatjet, the reconstructed dilepton mass and its $p_T$, the $\Delta \phi$ separation between the fatjet and the reconstructed dilepton system, the missing transverse energy, $\slashed{E}_T$, the mass of the Higgs fatjet and its transverse momentum, $p_T$ of the two $b$-tagged filtered subjets, the ratio of the $p_T$ of these $b$-tagged subjets and finally the rapidity of the reconstructed Higgs fatjet. During our training process, we do not require variables that are 100\% correlated but retain every other variable. Given that one of our final variables of interest is the reconstructed $Zh$ invariant mass, we refrain from using it as an input variable. For the BDT analysis, we use the \textsc{TMVA}~\cite{2007physics3039H} package in the root framework. During the analysis, we use 50\% of the samples for training and always ensure that there is no overtraining by requiring that the Kolmogorov-Smirnov statistic is at least $\mathcal{O}(0.1)$~\cite{KS}. After optimising the cut on the BDT variable, one finds that there are around 463 $q \bar{q} \to Zh$ (SM) and 820 $Zb\bar{b}$ events at 3 ab$^{-1}$, which amounts to the SM $qq\to Zh$ ($SM$) over rest of the background ($B$) ratio, $SM/B \sim 0.56$. Using the same training, we have respectively 44, 7 and 57 $Z+$ jets, $g g \to ZZ$ and $g g \to Z b \bar{b}$ backgrounds after the BDT cut. This yields $SM/B \sim 0.5$.

\subsection{The $W^{\pm}h$ channels}

For the $W^\pm h \to b \bar{b} \ell \nu$ analysis, we follow a very similar framework as before. The dominant backgrounds are the irreducible SM $W^{\pm}h$ and the reducible $W^{\pm}b\bar{b}$ channels. We also consider the fully and semi-leptonic $t\bar{t}$ events, $W^{\pm}+$ jets and $Z+$ jets, where $Z \to \ell^+ \ell^-$. The $W^{\pm}$ samples are generated at NLO QCD using the aforementioned method. The $W^{\pm}b\bar{b}$ samples are generated upon merging with an additional parton as described above. Unlike the $Zh$ channel, the $W^{\pm}h$ channel only has quark-initiated production mode. For the $Zh$ channel, it was quite simple to reduce the $t\bar{t}$ background by imposing a lower cut on $\slashed{E}_T$. For the $W^{\pm}$ study, the signal itself contains a final state with a neutrino and hence demanding a cut on $\slashed{E}_T$ will not only reduce the $t\bar{t}$ backgrounds but also a significant fraction of the signal. The signal samples are generated with $p_{T,h} > 150$ GeV and the invariant mass of the $Wh$ system, $m_{Wh} > 500$ GeV (we clarify this choice later). We use the same PDF choice as for the $Zh$ samples and the scales are chosen to be $\mu_F = \mu_R = m_{Wh}$. The backgrounds are generated with the same PDF choice at LO. The scales chosen for the background generation are $m_W$ for the $Wb\bar{b}$ and $W+$ jets samples and $2 m_t$ for the $t\bar{t}$ samples. Moreover, weak cuts are imposed on the background samples at the generation level. These include, $p_{T,(j,b)} > 15$ GeV, $p_{T,\ell} > 5$ GeV, $|y_{b/\ell}| < 3$, $|y_j| < 5$, $\Delta R_{b\bar{b}} > 0.1$, $\Delta R_{b\ell} > 0.2$ and 70 GeV $m_{b\bar{b}} < 155$ GeV. For the tree-level $W^+b\bar{b}, \; W^-b\bar{b}, \; t\bar{t}$, $W^+$+jets, $W^-$+jets and $Z+$ jets, we respectively use $K$-factor values of 2.68, 2.49, 1.35, 1.23, 1.18 and 1.13, computed within MCFM~\cite{Campbell:1999ah, Campbell:2011bn, Campbell:2015qma}. The $W^{\pm}b\bar{b}$ samples are generated upon merging with an additional parton, whereas the $W^{\pm}$+jets samples are merged with up to two additional partons. We separate the $Wh$ analysis into two parts depending on the charge of the isolated lepton. For the analysis, we require one isolated charged lepton. In contrast to the $Zh$ analysis, the $W^{\pm}h$ has a known ambiguity in the form of the $p_z$ component of the neutrino momentum. We deal with this by requiring that the invariant mass of the neutrino and the isolated lepton peaks around the $W$-boson mass. This gives us two solutions to $p_{z,\nu}$ and we demand that the solutions are always real. We discard events where complex solutions are encountered. We construct two invariant masses for the $Wh$ system for the two neutrino $p_z$ solutions, $m_{\text{fatjet}\ell\nu_{1,2}}$. Before implementing the BDT analysis, we employ certain loose cuts like $p_{T,\text{fatjet}} > 150$ GeV, $95 \; \text{GeV} < m_h < 155$ GeV, $m_{\text{fatjet}\ell\nu_{1,2}} > 500$ GeV and $\Delta R_{b_i,\ell} > 0.4$. On top of this we require certain number of fatjets, mass-drop and filtered subjets as discussed for the $Zh$ scenario. For the BDT analyses (one for $W^+h$ and another for $W^-h$), we train the samples upon considering the SM $Wh$ sample as the signal and the $Wb\bar{b}$, semi-leptonic and fully leptonic $t\bar{t}$ samples as backgrounds. Owing to multiple backgrounds, we impose relative weight factors to these backgrounds which are defined as $1/\mathcal{L}_{\textrm{gen}}$, where $\mathcal{L}_{\textrm{gen}}$ is the generated luminosity that depends on the production cross-section, including the $K$-factors, and the number of Monte Carlo generated events. Besides, NLO samples also contain negative weights for certain events, which we include while training the BDT samples. We also find that the effect of including the weight factor in our training is small, owing to the very small number of signal events having negative weights (less than 4\% percent). We optimise the BDT analysis for $W^+h \; (W^-h)$ and find 1326 (901) events for the signal and 4473 (3476) $W^+b\bar{b} \; (W^-b\bar{b})$ events at 3 ab$^{-1}$. The number of surviving events for $t\bar{t}$, $W+$ jets and $Z+$ jets are much smaller. Ultimately, we find $SM/B \sim 0.28 \; (0.24)$ for $W^+h \; (W^-h)$.


\section{Analysis and Results}
\label{analysis}
In this section we describe how we  obtain our final sensitivity estimates and present our main results. We will consider only the interference contribution in this study which in any case is expected to be dominant piece below the EFT cut-off. There is no conceptual hurdle in including also the squared terms, as \eq{sumsquare} is still equally valid, and the reasons for neglecting them are only practical. We first consider the contact terms, $g^h_{Vf}$, which can be very precisely constrained in the high energy bins. Once  these couplings are very precisely constrained we will turn to the lower energy bins where there are a sufficient number of events to carry out an angular moment analysis to constrain the other couplings. All the results we will  present in this section will be for an integrated luminosity of 3 ab$^{-1}$.

\subsection{Bounds on contact terms}

As already discussed, at high energies the EFT deviations are dominated by the contribution of the contact interactions, $g^h_{Vf}$, to $a_{LL}$. Because this contribution grows quadratically with energy relative to the SM $Vh$ contribution,  it can be very precisely constrained by probing high energy bins. Unfortunately some of the bins providing maximum sensitivity have too few events for an angular moment analysis. We thus constrain these couplings simply using the final state invariant mass distribution. Following Ref.~\cite{Banerjee:2018bio}, where this procedure was carried out for the $Zh$ mode, we construct a bin-by-bin $\chi^2$ function assuming the expected number of events is given by the SM and the observed by the SMEFT. To ensure that we do not violate EFT validity we neglect any event with a final state invariant mass above the cut-off, which is evaluated for a given value of the anomalous couplings, by setting the Wilson coefficients in \eq{wilson} to unity. For an integrated luminosity of 3 ab$^{-1}$, we obtain the sub-per-mille level bounds at the one sigma level, \footnote{Note the small difference in the bound on $g^h_{Z\textbf{p}}$, compared to the one obtained in Ref.~\cite{Banerjee:2018bio} because of a more rigorous inclusion of NLO effects and other variations in the analysis strategy.}
\bea
|g^h_{WQ}| &<& 6\times 10^{-4}\label{contactw}\nn\\
|g^h_{Z\textbf{p}}| &<& 4\times 10^{-4}.\label{contactz}\nn\\
\eea

 \subsection{Angular Moment analysis}
 
 Now that $g^h_{WQ}$ and $g^h_{Z\textbf{p}}$ are strongly constrained from the higher energy bins, we turn to the lower energy bins with enough events to perform an angular moment analysis to constrain the other couplings. Ideally we should marginalise over the effect of contact terms also in the lower bins, but as we will see the expected bounds on the contact terms are almost two orders of magnitude smaller than that of the other couplings, and thus their effect is negligible in the lower energy bins. Therefore we will ignore them in further analysis. We first split  our  simulated  events into 200 GeV  bins of the final state invariant mass. To obtain  the angular moments we first  convolute  the events in each energy bin with  the weight functions using \eq{angsum}. As the $CP$-even and odd couplings contribute to a mutually exclusive set of angular moments we construct two separate  bin-by-bin $\chi^2$ functions as follows,
\bea
\label{2chis}
\chi^2(\delta \hat{g}^h_{VV}, {\kappa}^\textbf{p}_{VV})&=& \sum_{ij} \frac{\left(a_i^{EFT} (M_j)- a^{SM}_i (M_j)\right)^2}{(\Sigma_i (M_j))^2}\nn\\
\tilde{\chi}^2( {\tilde{\kappa}}^\textbf{p}_{VV})&=& \sum_{ij} \frac{\left(\tilde{a}_i^{EFT} (M_j)- \tilde{a}^{SM}_i (M_j)\right)^2}{(\Sigma_{i} (M_j))^2}
\eea
where $\kappa^\textbf{p}_{VV}, \tilde{\kappa}^\textbf{p}_{VV}$ are the same as $\kappa_{WW}, \tilde{\kappa}_{WW}$ for $V=W$ and defined in \eq{kappap} for $V=Z$.  In the above equation, we include only  the $CP$-even ($CP$-odd) angular moments in $\chi^2$ ($\tilde{\chi}^2$),  the index $i$ indicates the different moments and $M_j$ labels   the invariant mass bins. The squared error in the denominator  is computed using \eq{angerr} on the background sample (which includes the SM $Vh$ contribution) where $\hat{N}$ in this case is the total number of background events in the $j$-th bin.

Once again the contributions due to  $\kappa^\textbf{p}_{VV}$ and $\tilde{\kappa}^\textbf{p}_{VV}$    grow with energy and one must be careful about EFT validity. For a given value of the coupling we estimate the cut-off $\Lambda$ using \eq{wilson} putting the all the Wilson coefficients  to unity. We ignore any event that has final state invariant mass above 1500 GeV, a value smaller than the cut-off corresponding to the size of the couplings we will eventually constrain. The most sensitive bins for the analysis of the contact term, on the other hand, are bins higher than 1500 GeV.  The contribution due to $\hat{g}^h_{VV}$ does not grow with energy with respect to the SM and thus the bounds on this coupling are in any case dominated by the contribution  from the lowest energy bins in our analysis.  

We now discuss the results for  the $Zh$ and $W^\pm h$  modes separately before presenting our combined bounds. The individual bounds are important as they do not assume \eq{relation} which has been derived assuming that electroweak symmetry is linearly realised. In fact, the independent measurement of couplings involving the $Z$ and $W$  can be used to verify \eq{relation}  as a prediction of linearly realised electroweak symmetry.

 \begin{figure}[!t]
\begin{center}
  \subfigure[]{\includegraphics[width=0.46\textwidth]{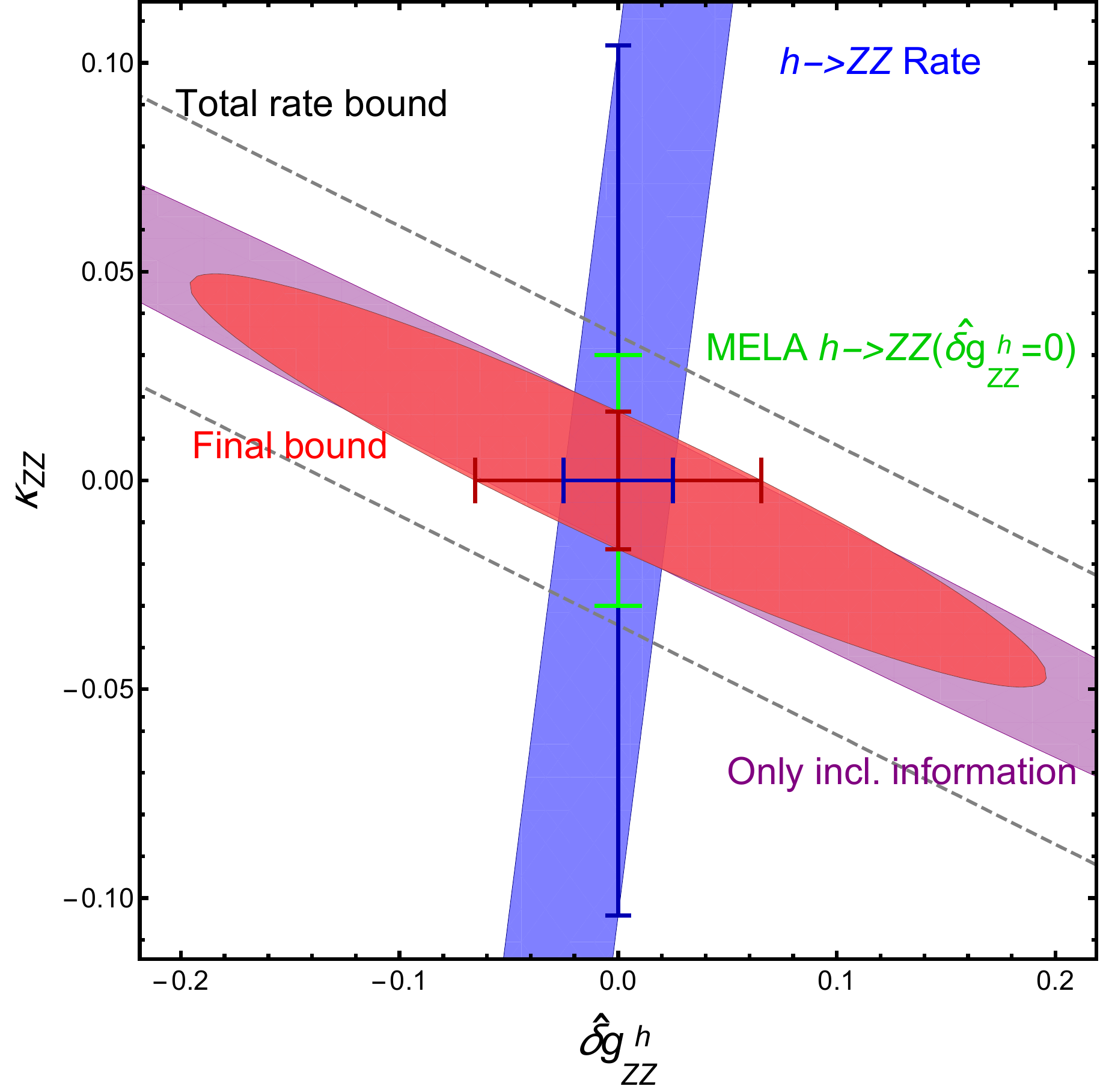} \label{boundsz}}
  \subfigure[]{\includegraphics[width=0.45\textwidth]{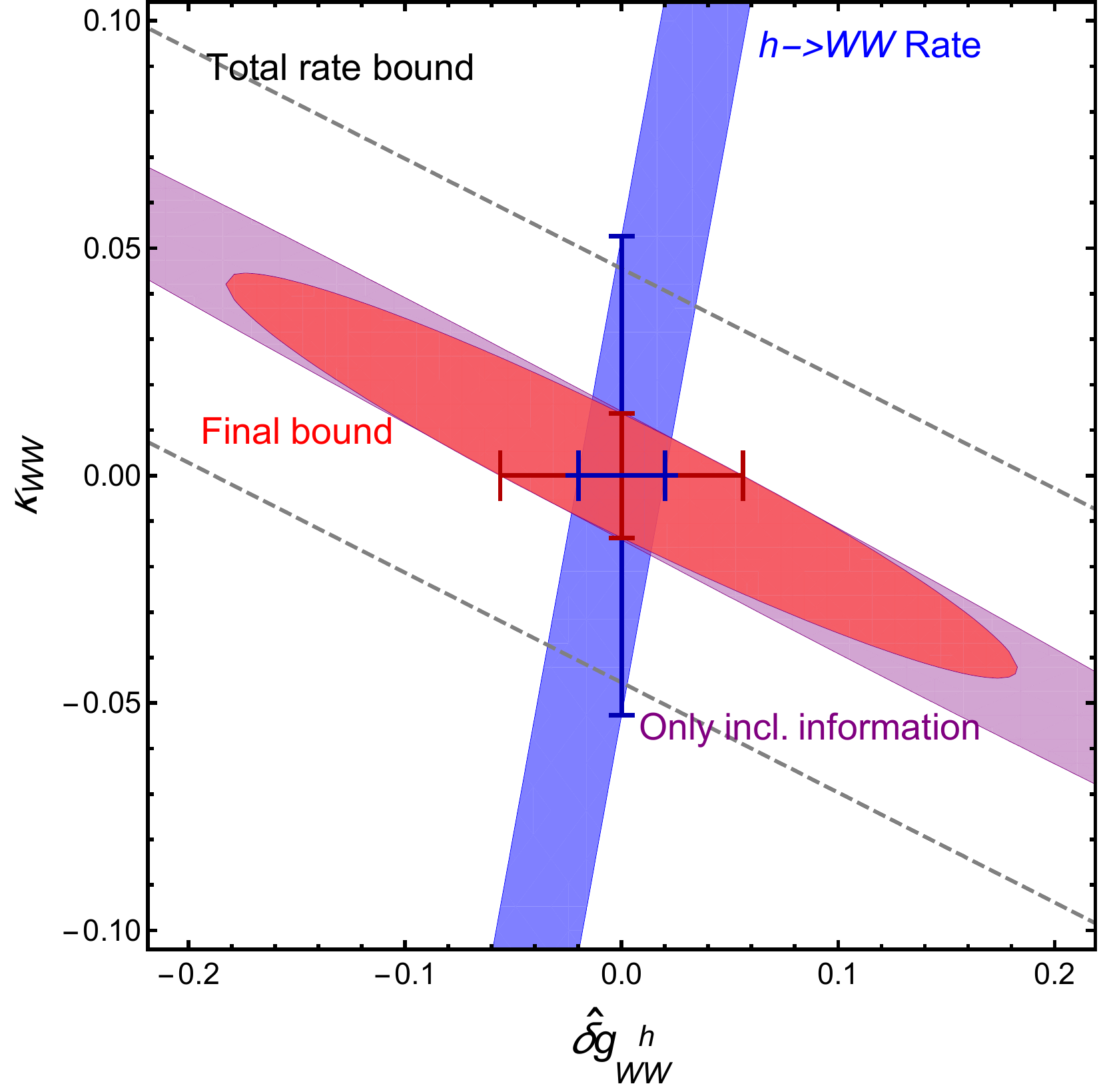} \label{boundsw}}
 \caption{(a) Bounds at 65$\%$ CL on the $CP$-even anomalous couplings from $Zh$ production with 3 ab$^{-1}$ integrated luminosity, assuming that the contact term has been very precisely constrained (see \eq{contactz}). We show the improvement of the bounds as more and more differential information is included in the fit. The dashed lines show the bound just from the total rate. The purple region includes differential information at the level of the $Z$-boson four momentum such as the final state invariant mass distribution and $\Theta$-distribution. Finally the red region includes information from all the angular moments including the cross-helicity interference terms. The blue band shows the bound from $h \to ZZ \to 4\ell$ rate using the results of Ref.~\cite{Cepeda:2019klc}. The bars show the bounds on one of the couplings when the other coupling is 0. The green bar shows the bound obtained using the Matrix Element Likelihood Analysis (MELA) in Ref.~\cite{Anderson:2013afp} and assuming $\delta\hat{g}^h_{ZZ}=0$. (b) Same as in (a) but for the $W^\pm h$ mode where there is no bound from MELA.}
 \end{center}
\end{figure}

\label{results}
\subsubsection{$Zh$ mode}

The bound obtained for the two $CP$-even couplings is shown in Fig.~\ref{boundsz}.  To show the power of our method we show the progression of the bounds obtained as the differential information used is gradually increased. The bound obtained, if one uses only total rate to constrain a linear combination of the two couplings, $\delta \hat{g}^h_{ZZ}$ and  ${\kappa}^\textbf{p}_{ZZ}$ is shown by the two dashed lines. Next we include distributions of the final state invariant mass and other differential  information at the level of $Z$-boson four momentum, \textit{i.e.}, the decay products of the $Z$-boson are treated inclusively, and obtain the excluded region shown in purple; for this we  include only the angular moments $\hat{a}_{1}$ and  $\hat{a}_{3}$, extracted using the weights in Sec.~\ref{basic}, thus using  information of the $\Theta$-distribution. The analysis at this stage is comparable to a regular SMEFT analysis that includes a few standard differential distributions. Finally to obtain our final bound shown in red we include in \eq{2chis},  the moments $\hat{a}'_{1}, \hat{a}'_{3}, a^2_{LT}$ and $a_{TT'}$  in $\chi^2$ (see Sec.~\ref{zh}).  Recall that $\hat{a}'_{1}$ and $\hat{a}'_{3}$ are linear combinations of the original angular moments $a_{LL}$ and $a^2_{TT}$ defined in Sec.~\ref{alternate}. The main improvement in sensitivity  in the final bounds comes from $a^2_{LT}$  the effect of which can be captured only by   a careful study of the joint $(\Theta, \theta, \varphi)$ distribution as pointed out in Ref.~\cite{Banerjee:2019pks}. While this is clearly something beyond the scope of a regular cut-based analysis, as one would need to take into account all the correlations of the final state phase space, the angular moment approach captures it effortlessly.

We show also the projected  bounds from the $h\to ZZ\to 4\ell$ process in Fig.~\ref{boundsz}.  The blue band shows the bound from the $h\to ZZ\to 4\ell$ rate whereas the green bar is the bound obtained using the Matrix Element Likelihood Analysis (MELA) framework~\cite{Anderson:2013afp}. As far as $\kappa^\textbf{p}_{ZZ}$ is concerned, we see  that the bound obtained  from $Zh$ production  using our methods surpass the other existing projections shown  in Fig.~\ref{boundsz}~\footnote{A bound using the matrix element method for $pp \to Zh$ may potentially match our bounds but the results in Ref.~\cite{Anderson:2013afp} are unfortunately not comparable to ours  as these studies include high energy phase space regions where the EFT contribution is many times that of the SM. The methodology iused to obtain these bounds, thus, violate our assumption of ${\cal O}(1)$ Wilson coefficients.}. In the horizontal direction our bounds might seem redundant once the
 $h\to ZZ\to 4\ell$ process is taken into account, but if one allows for $hbb$ coupling deviations our bounds become the measurement of a truly independent effect, see \eq{rescale}. 
 
 The CP odd coupling, $\tilde{\kappa}^\textbf{p}_{ZZ}$ is constrained using the function $\tilde{\chi}^2$ in \eq{2chis} which includes the moments  $\tilde{a}^1_{LT}$ and $\tilde{a}_{TT'}$. We finally obtain the one sigma level bound,
\bea
\label{cpoddz}
|\tilde{\kappa}^\textbf{p}_{ZZ}|<0.03.
\eea

\subsubsection{$W^{\pm}h$ modes}
 We show the progression of the bounds for the $CP$-even case at different stages of inclusion of differential information in Fig.~\ref{boundsw}. The dashed lines show bounds from the total rate and the purple region shows the bound obtained by  including only the angular moments, $a_{LL}$ and $a^2_{TT}$, using the weights  in Sec.~\ref{basic},  that encapsulate the differential information at the level of the $Z$-boson treating its decay products inclusively.  For our final bound  in the $CP$-even case shown in red  we include the effect of all the relevant angular moments for this case, namely,   $a_{LL}, a^2_{TT}$ and $a_{TT'}$ (see Sec.~\ref{wh}) where for the first two moments we extract the linear combinations $\hat{a}'_{1}$ and $\hat{a}'_{3}$ described in Sec.~\ref{alternate}. We show also the projected bounds from the $h \to WW \to2 l 2\nu$ decay rate  in blue to which our bounds are complementary (recall again that, what our bounds actually probe is a linear combination also involving  $hb\bar{b}$ coupling deviations, see \eq{rescale}). In this case there is no competing bound on $\kappa_{WW}$ from the  $h \to WW$ mode presumably because the neutrinos in the final state make much of the differential information inaccessible in this case. Thus our bounds on $\kappa_{WW}$ from the $pp \to W^\pm h$ process is likely to be the best bound on this coupling possible.

Again the CP odd coupling, $\tilde{\kappa}_{WW}$ is constrained by  including the moment  $\tilde{a}_{LT1}$ in the function $\tilde{\chi}^2$ in \eq{2chis}. We finally obtain the one sigma level bound,
\bea
\label{cpoddw}
|\tilde{\kappa}_{WW}|<0.04.
\eea

\begin{figure*}[!t]
\begin{center}
  \includegraphics[width=0.6\textwidth]{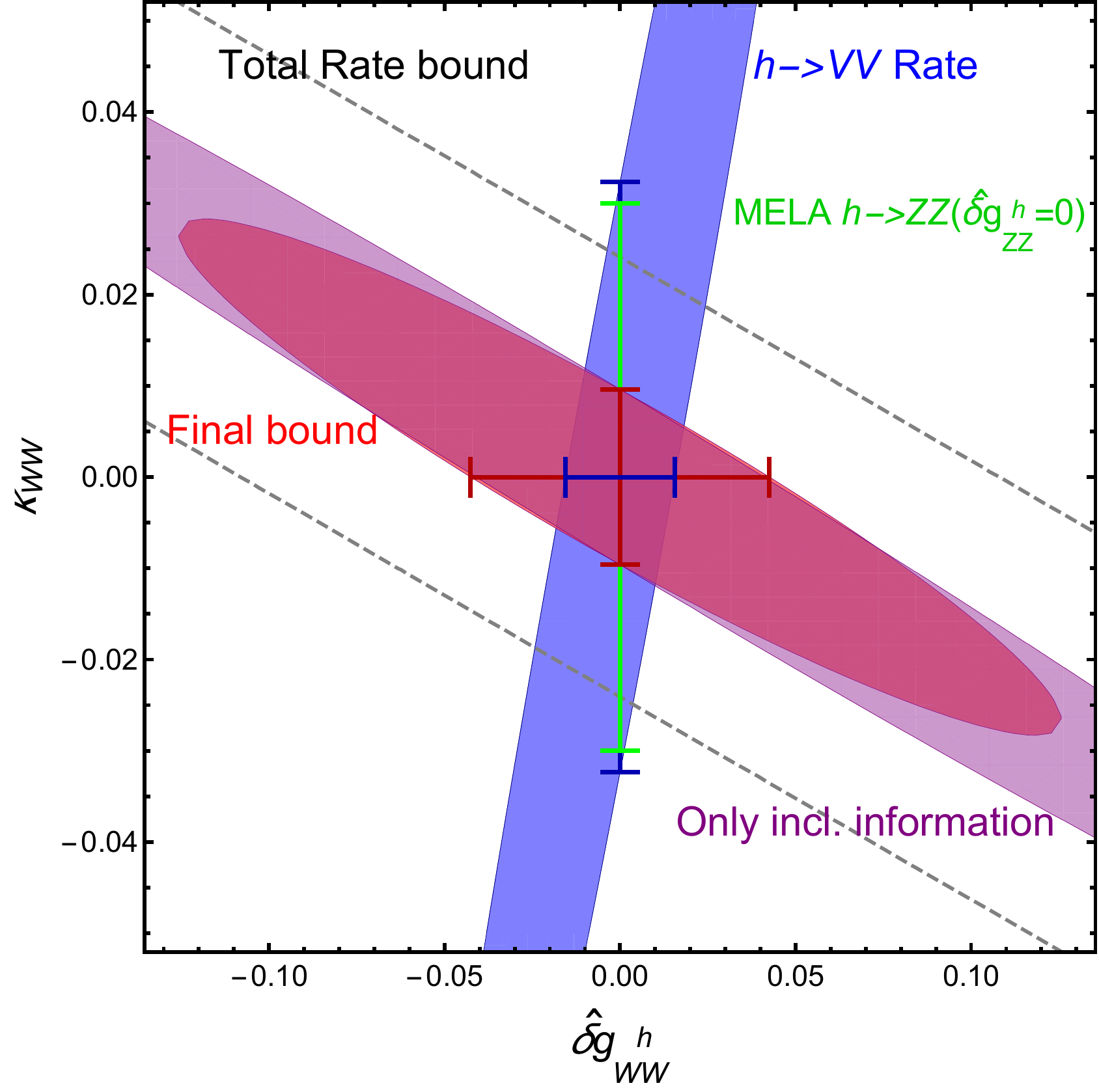} 
 \caption{Bounds at 65$\%$ CL on the $CP$-even anomalous couplings, with 3 ab$^{-1}$ integrated luminosity,  after combining results from  $Zh$ and $Wh$ production using \eq{relation} and  assuming that the contact terms have been very precisely constrained (see \eq{contactz}). Again, we show the progression of the bounds as more and more differential information is included in the fit. The dashed lines show the bound just from the total rate in both processes. The purple region includes differential information at the level of the $Z/W$-boson four momentum. The red region is our final bound and includes information from all the angular moments. The blue band shows the bound from a combination of $h \to WW\to 2l 2\nu$  and $h \to ZZ\to 4\ell$ rate  using the results of Ref.~\cite{Cepeda:2019klc}. The bars show the bounds on one of the couplings when the other coupling is zero. The green bar shows the bound implied by the bound on $\kappa_{ZZ}$ using the Matrix Element Likelihood Analysis (MELA) in Ref.~\cite{Anderson:2013afp} and assuming $\delta\hat{g}^h_{ZZ}=0$.}
  \label{combo}
\end{center}
\end{figure*}

We see that we obtain bounds of similar size from the $pp \to Wh$ and $pp\to Zh$ processes on the respective anomalous couplings.  The fact that the couplings can be independently measured is very important as we can then use these measurements to test the correlations in \eq{relation} which  in turn tests whether electroweak symmetry is linearly realised or not. An alternative approach would be to use the correlation to combine the bounds from $Wh$ and $Zh$ production as we show in the next subsection.

\subsubsection{Combination of $Zh$ and $W^{\pm}h$ modes}

In Fig.~\ref{combo} we show the bounds obtained after combining the results of using \eq{relation}, thus assuming electroweak symmetry is linearly realised.  Again, we show the bound obtained at various levels of inclusion  of differential data. The dashed lines show the bound just from the total rate, the purple region includes differential information at the level of the $Z/W$-boson four momentum and the red region is our final bound including all angular moments. The blue band shows the bound from a combination of $h \to WW\to 2l 2\nu$  and $h \to ZZ\to 4\ell$ rate. The green bar shows the MELA bound from   Ref.~\cite{Anderson:2013afp} on $\kappa_{ZZ}$ assuming $\delta\hat{g}^h_{ZZ}=0$, translated to this plane.

\subsubsection{Comparison with  bounds from $WZ$ and $WW$  production}
\label{tgc}

If electroweak symmetry is linearly realised  bounds on $\kappa_{WW}$ and $g^h_{WQ}$ can be extracted also from double gauge boson production using \eq{tgc1} and \eq{tgc2}. For instance $WZ$ production  at high energies  constrains precisely the linear combination of $Z$-pole couplings and TGCs that appears in the right hand side of \eq{tgc1} at the sub per-mille level~\cite{Franceschini:2017xkh}. This bound is of the same size as the one obtained in \eq{contactw} in  this work. Combining the two bounds will thus yield a significantly improved bound compared to the individual ones.  This is also true for \eq{tgc2} where the least constrained coupling in the right hand side, $\delta \kappa_\gamma$, can be bounded at the level  of a few percent in $WW$ production~\cite{Grojean:2018dqj}; this is comparable to our bound on $\kappa_{WW}$ in Fig.~\ref{boundsw} and Fig.~\ref{combo} once we marginalise over $\delta \hat{g}^h_{WW}$. In making the last statement we used the fact that $Z$ couplings to quarks that  appear in the right hand side of \eq{tgc2} and also affect $WW$ production are measured more precisely at the per-mille level~\cite{Falkowski:2014tna}.  

 Alternatively, the fact that the left and right hand sides of \eq{tgc1} and \eq{tgc2}  can be measured with similar precision, in double gauge boson  and Higgs-strahlung processes, means that one can actually verify \eq{tgc1} as a test of linearly realised electroweak symmetry at the HL-LHC.

\section{Conclusions}
\label{conclusions}

The precise measurement of Higgs boson properties will be one of the legacies of the LHC's scientific achievements. Potential deformations of the Higgs boson's couplings to other particles compared to Standard Model predictions can be cast into limits on Wilson coefficients of effective operators originating in the SMEFT framework. To obtain predictive limits on the highly complex system of SMEFT operators, it is necessary to measure Higgs interactions in various production and decay channels. One of the most important ones to establish the nature of the Higgs boson and its embedding into the scalar sector are its couplings to massive gauge bosons, \textit{i.e.}, the $W$ and $Z$ bosons.

We proposed a novel method to probe the full structure of the Higgs-gauge boson interactions in Higgs-associated production. Using the helicity amplitude formalism and expanding the squared matrix elements into angular moments the whole process can be expressed in terms of nine trigonometric functions. This is true not only in the SM but also in the D6 SMEFT.  Extracting the coefficients of these functions, the so called angular moments, is a powerful and predictive way of encapsulating the full differential information of this process. As differential information can encode signatures of EFT operators in subtle ways, maximally mining the differential information is essential to obtain the best possible bounds on the EFT operators. As the actual interpretation of the measurement relies now on a shape analysis of a small number of trigonometric functions, strong constraints can be obtained, provided experiments are going to publicise measurements of these functions. Thus, we encourage the experimental collaborations to provide such measurements for various Higgs production processes\footnote{The provision of measurements of the moments and basis functions will allow for an ideal approach to perform hypothesis testing for effective operators. As such it will improve on current initiatives of using so-called simplified cross section measurements \cite{Berger:2019wnu} in global fits.}.

The efficacy of this method relies crucially on whether the theoretical form of the original angular distribution can be preserved despite effects like experimental cuts, showering and hadronisation. In this article, we carried out a detailed collider simulation of the Higgs-strahlung process, including these effects,  before applying  the method of angular moments. The results we find are encouraging, indicating that a shape analysis using the trigonometric basis functions can set the most sensitive limits on effective operators within the SMEFT framework. While the high energy behaviour of the process results in the strongest possible bounds on the $hVff$ contact terms (see \eq{contactz}), the full angular moment analysis  leads to the strongest reported  bounds on the $h V_{\mu \nu}V^{\mu \nu}$ (see Figs.~\ref{boundsz},~\ref{boundsw} and \ref{combo}) and  $h V_{\mu \nu}\tilde{V}^{\mu \nu}$ (see \eq{cpoddz} and \eq{cpoddw}).

We aim to extend this method to various other Higgs/electroweak production and decay processes such as weak boson fusion~\cite{jeppe}, the $h \to ZZ \to 4\ell$ decay~\cite{oscar} and diboson production~\cite{dibosonmom}. One can then perform a full global fit including this enlarged set of observables to obtain the best possible bounds on the SMEFT lagrangian. 

\section*{Acknowledgements}
RSG would like to  thank Amol Dighe for  pointing out that  the method of moments may be very suitable for differential  SMEFT analyses of this kind. We would also like to thank Shilpi Jain and Marek Sch\"{o}nherr for helpful discussions. S.B. was supported by a Durham Junior Research Fellowship COFUNDed by Durham University and the
European Union, under grant agreement number 609412.

\bibliographystyle{JHEP}
\bibliography{references}    
\end{document}